\documentclass[useAMS, usenatbib]{mn2e}
\usepackage{graphicx}
\usepackage{mathptmx}

\newcommand{\msun}{M$_\odot$}
\newcommand{\mjybeam}{mJy\,beam$^{-1}$}
\newcommand{\clfind}{{\sc clfind}}
\newcommand{\gclumps}{{\sc gaussclumps}}
\newcommand{\cupid}{{\sc cupid}}
\newcommand{\apj}{ApJ}
\newcommand{\apjl}{ApJ}

\newcommand{\apjss}{ApJS}
\newcommand{\aap}{A\&A}

\newcommand{\mnras}{MNRAS}

\newcommand{\araa}{ARA\&A}

\title[The properties of SCUBA cores in the Perseus molecular cloud]{The properties of SCUBA cores in the Perseus molecular cloud: the bias of
  clump-finding algorithms}
\author[E.~I.~Curtis and J.~S.~Richer]{Emily~I.~Curtis$^{1\rmn{,}2}$\thanks{E-mail:
e.curtis@mrao.cam.ac.uk} and John~S.~Richer$^{1\rmn{,}2}$\\
$^{1}$Astrophysics Group, Cavendish Laboratory, J. J. Thomson
Avenue, Cambridge, CB3 0HE\\
$^{2}$Kavli Institute for Cosmology, c/o Institute of Astronomy, University
of Cambridge, Madingley Road, Cambridge, CB3 0HA}

\begin{document}

\date{Accepted 2009 October 21; Received 2009 October 7; in original
  form 2009 August 11}

\pagerange{\pageref{firstpage}--\pageref{lastpage}} \pubyear{2009}

\maketitle

\label{firstpage}

\begin{abstract}

We present a new analysis of the properties of star-forming cores in
the Perseus molecular cloud, identified in SCUBA 850\,\micron\
data originally presented by \citet{hatchell05}. Our goal is to
determine which core properties can be robustly identified and which depend on the extraction technique. Four regions in the cloud are examined: NGC~1333, IC348/HH211, L1448 and
L1455. We identify clumps of dust emission
using two popular automated algorithms, \clfind\ and \gclumps, finding
85 and 122 clumps in total respectively. Using the catalogues of
\citet{hatchell07a}, we separate these clumps into starless, Class 0 and
Class I cores. Some trends are true for both populations: clumps become
increasingly elongated over time; clumps are consistent with constant
surface brightness objects (i.e.\ $M\propto R^2$), with an average
brightness $\approx 4$--10 times larger than the surrounding molecular
cloud; the clump mass distribution (CMD) resembles the stellar intial
mass function, with
a slope $\alpha = -2.0 \pm 0.1$ for \clfind\ and $\alpha = -3.15 \pm
0.08$ for \gclumps, which straddle the Salpeter value ($\alpha = -2.35$). The mass at
which the slope shallows (similar for both algorithms at $M\approx 6$\,M$_\odot$)
implies a star-forming efficiency of between 10 and 20\,per
cent. Other trends reported elsewhere depend critically on the
clump-finding technique: we find protostellar clumps are both smaller (for
\gclumps) and larger (for \clfind) than their starless counterparts;
the functional form, best-fitting to the CMD, is different for the two
algorithms. The \gclumps\ CMD is best-fitted with a log-normal
distribution, whereas a broken power law is best for \clfind; the
reported lack of massive starless cores in previous studies (e.g.\
\citealt{hatchell07a,hatchell08}) can be seen in the \clfind\ but not
the \gclumps\ data. Our approach, exploiting two extraction techniques, highlights similarities
and differences between the clump populations, illustrating the
caution that must be exercised when comparing results from different
studies and interpreting the properties of samples of continuum cores. 
\end{abstract}


\section{Introduction} 

Stars form out of gas in the densest areas of molecular
clouds. Early studies of molecular cloud structure (e.g.\ \citealp{blitz80})
concluded that a division into discrete clumps of emission was the best
description. Within such clumps, star formation occurs inside denser
cores \citep{myers83b}. A growing consensus now maintains that
molecular clouds have a scale-free structure governed by turbulence (e.g.\
\citealp{elmegreen96,stutzki98,elmegreen04}), with clumps only an arbitrary
categorization. We may reconcile these viewpoints to some extent; the
clump population has a power-law spectrum of mass implying the overall
cloud has a similar distribution (e.g.\ \citealp*{williams00}). Equally, the self-similarity of molecular
clouds must break down where gravitational collapse
becomes important. 

This paper focusses on the decomposition of molecular clouds
into \emph{clumps} of emission. The utility of such a description depends on
whether the located clumps accurately represent star-forming
cores. Clumps are typically located in either dust continuum or
spectral-line datasets, which both have their own advantages and drawbacks. Continuum observations select
high-density cores, which tend to be self-gravitating, but associate more mass to
clumps than exists in reality, since objects may be superposed along the
line of sight (e.g.\ \citealp*{smith_clark08}). Clump-finding analyses on spectral-line cubes, with their
added velocity dimension, should enable clumps to be more accurately
assigned but are subject to larger uncertainties from the details of
the radiative
transfer than continuum emission. Ultimately, high signal-to-noise
ratios are required in both cases for clumps to converge towards the underlying core
population \citep{ballesteros02}.  

Nevertheless, many insights into star formation from clump populations remain compelling. \citet*{motte98} first pointed
out that the mass distribution of compact continuum clumps (the clump
mass distribution; hereafter CMD) is remarkably similar to the initial
mass function of stars (IMF). At high masses, the slope is consistent with a Salpeter power law
($\mathrm{d}N/\mathrm{d}M\propto M^{-2.35}$, \citealp{salpeter55}), considerably steeper than the power law
for CO clumps  ($\mathrm{d}N/\mathrm{d}M\propto M^{-1.6}$,
e.g.\ \citealp{blitz93}). At lower masses, the CMD slope becomes
shallower, around the point that samples start to suffer from
incompleteness. This suggests that the mass of a star is established
at its earliest phases (see also \citealp*{alves07}) and would seem to rule out models where the
shape of the IMF is formed later, through e.g.\ dynamical interactions
\citep{bate05}. However, the mapping of the CMD on to the
IMF is not straightforward and many evolutionary schemes would fit the
present data (e.g.\ \citealp{swift08}).

Many wide-field surveys are about to or have just begun, which
will locate and characterize thousands of star-forming cores from
continuum data, e.g.\ the
James Clerk Maxwell Telescope (JCMT) and \emph{Herschel} Gould Belt Surveys \citep{gbs,andre05}. One
of the key science drivers for these projects is the measurement of
the CMD to high precision at low masses, into the brown-dwarf regime
($\la 0.08$\,\msun). Thus, an examination
of different source-finding techniques is particularly timely. Many
different methods have been deployed in the literature to locate
clumps in molecular-line and continuum data, from identifications by
eye to automated algorithms. However, few studies \emph{compare} the
sources located with different techniques (a notable exception is \citealt{schneider04}). In this paper, we
closely compare the populations of continuum clumps found in the
Perseus molecular cloud (hereafter simply Perseus) using the two most popular
automated algorithms, \clfind\ \citep*{williams94} and \gclumps\
\citep{stutzki90}, to highlight their differences and biases. We
re-examine SCUBA 850\,\micron\ data, presented originally by
\citet{hatchell05}, in the four clusters of cores where we have complementary
HARP spectral-line data \citep*{paper1}, namely: NGC~1333,
IC348/HH211 (IC348 for short),
L1448 and L1455. Although only investigating a sub-set of the SCUBA
data limits our sample size, these are the sources whose kinematics we
will investigate subsequently and any CMDs from the spectral-line data
will be directly comparable (Curtis \& Richer, in prep.). The regions
selected still encompass the majority of the SCUBA cores identified by
\citet{hatchell05,hatchell07a}, 58 out of 103; so we can extrapolate
any conclusions to the entire cloud population. This work is divided
into two principal parts. In Section \ref{sec:clumppopulation}, we identify clumps using the two
algorithms before matching them to a  catalogue of SCUBA cores, classified as protostellar or starless on the basis of
their SEDs by \citet{hatchell07a} (hereafter
\defcitealias{hatchell07a}{H07}\citetalias{hatchell07a}).
Second, Sections \ref{sec:properties} and \ref{sec:cmd} present an analysis of the physical properties
of the cores before we summarize our conclusions in Section \ref{sec:summary}.

\subsection{Nomenclature} 

Many different terminologies have been used to
describe the hierarchical structure in molecular clouds; we follow \citet{williams00}. Within molecular
clouds, individual over-densities are termed \emph{clumps}, these are
the objects identified by automated algorithms and do not necessarily
go on to form stars. Clumps may contain \emph{cores}, which
are the direct precursors of individual or multiple stars. Every clump
that does not contain an embedded object is referred to as
\emph{starless}. Of these starless cores, only a subset, the
\emph{prestellar} cores (formerly pre-protostellar cores, \citealt{wardthompson94})  will be
gravitationally bound and thus go on to form stars. 

\section{Observational data}

We extracted fully calibrated and reduced SCUBA 850\,\micron\ maps
across the four regions in Perseus we observed with HARP from the data presented by \citet{hatchell05}, where we refer
the reader for details of the observations and processing. In short, the data were taken during 20 nights between 1999 and
  2003. The sky opacity measured at 225\,GHz varied from $\tau_{225}=0.039$ to
  0.080, with most of the data taken in good conditions
  ($\tau_{225}\approx 0.05$). The beam size is 14\,arcsec (0.017\,pc at 250\,pc, our assumed distance to Perseus) and
the maps are sampled on a 3\,arcsec grid. The rms noise level varies across the
  maps from typical values of $\sigma_\rmn{rms}=26$\,\mjybeam\ in
  NGC~1333 and IC348 to higher values, 32 and 46\,\mjybeam\ in L1448
  and L1455. Hereafter, to be consistent with the
  \citeauthor{hatchell05} studies and therefore allowing a direct
  comparison of our source catalogues, we take the noise to be
  $\sigma_\rmn{rms}=35$\,\mjybeam, the typical value for the \emph{entire} cloud. 

\section{The population of dust clumps} \label{sec:clumppopulation}

(Sub)millimetre continuum mapping selects young, cold cores (5--20\,K)
at the earliest stages of star formation. These largely
self-gravitating cores are intermediate in properties between CO
clumps and infrared young stellar objects (see
\citealp{wardthompson07} for a review). The dust's thermal emission is
optically thin nearly everywhere at these wavelengths, accurately
tracing density throughout the cloud. 

\subsection{Algorithms}

We use \cupid\ \citep{berry07}\footnote{Distributed as part of the
Starlink software collection, see
\texttt{http://starlink.jach.hawaii.edu}.} to identify clumps of
emission in the SCUBA 850\,\micron\ maps, exploiting the two most widely
used algorithms: 
\begin{enumerate}
\item `Clumpfind' or \clfind\ \citep{williams94}, which contours the
  data at evenly-spaced levels, starting close to the peak value. All
  distinct contiguous areas are considered above each contour level in
  turn. If an area has no pixels previously assigned to a clump, it is
  marked as a new clump. Alternatively if there is a pre-existing
  clump assignment, that clump is extended to the next lowest contour
  level. When two or more different assignments exist in a area, the
  pixels are distributed using a friend-of-friends algorithm.
\item \gclumps\ \citep{stutzki90}, which uses a least-squares fitting
  procedure to break-down the data into a number of Gaussian-shaped
  clumps, in a fashion similar to the {\sc clean} algorithms used in
  interferometric mapping.  
\end{enumerate} \clfind\ is a partitioning
method with each pixel only allocated to a single clump, whereas
\gclumps\ permits any number of clumps to overlap at a
single spatial position. This means \gclumps\ is better at
handling blended sources, although its output clumps
necessarily conform to a strict Gaussian profile as opposed to the
arbitrary shapes from \clfind. 

The way each algorithm find clumps of emission is controlled
  by a number of tuning parameters. Our overall aim in this paper is
  to compare the populations of clumps found \emph{independently} by
  \clfind\ and \gclumps. We therefore try to find the same objects
  with both algorithms, namely clumps larger than the beamsize with
  peaks greater than $4 \sigma_\rmn{rms}$, but do not try to make the
  algorithms find exactly the same identifications clump-to-clump. In so
  doing, we follow the majority of previous studies and select
  parameters for each algorithm that are considered optimal for
  rejecting spurious detections. In an alternative approach, one could
  try to produce clump catalogues which were as similar as
  possible. However, the standard algorithm outputs we supply are
  quite different because of the very different way clumps are
  found with the two algorithms. Given that we can never
  definitively know the true, underlying clump population, then selecting
  one algorithm population to make more similar to the other
  would be completely arbitrary \emph{a priori}. Indeed, it is only
  through detailed analysis, such as the one presented in this paper,
  or Monte-Carlo simulations that such a
  decision regarding algorithm selection could be made.

\subsubsection{\clfind} 

\begin{figure*}
\begin{center}
\begin{minipage}{0.48\textwidth}
\includegraphics[height=\textwidth,angle=270]{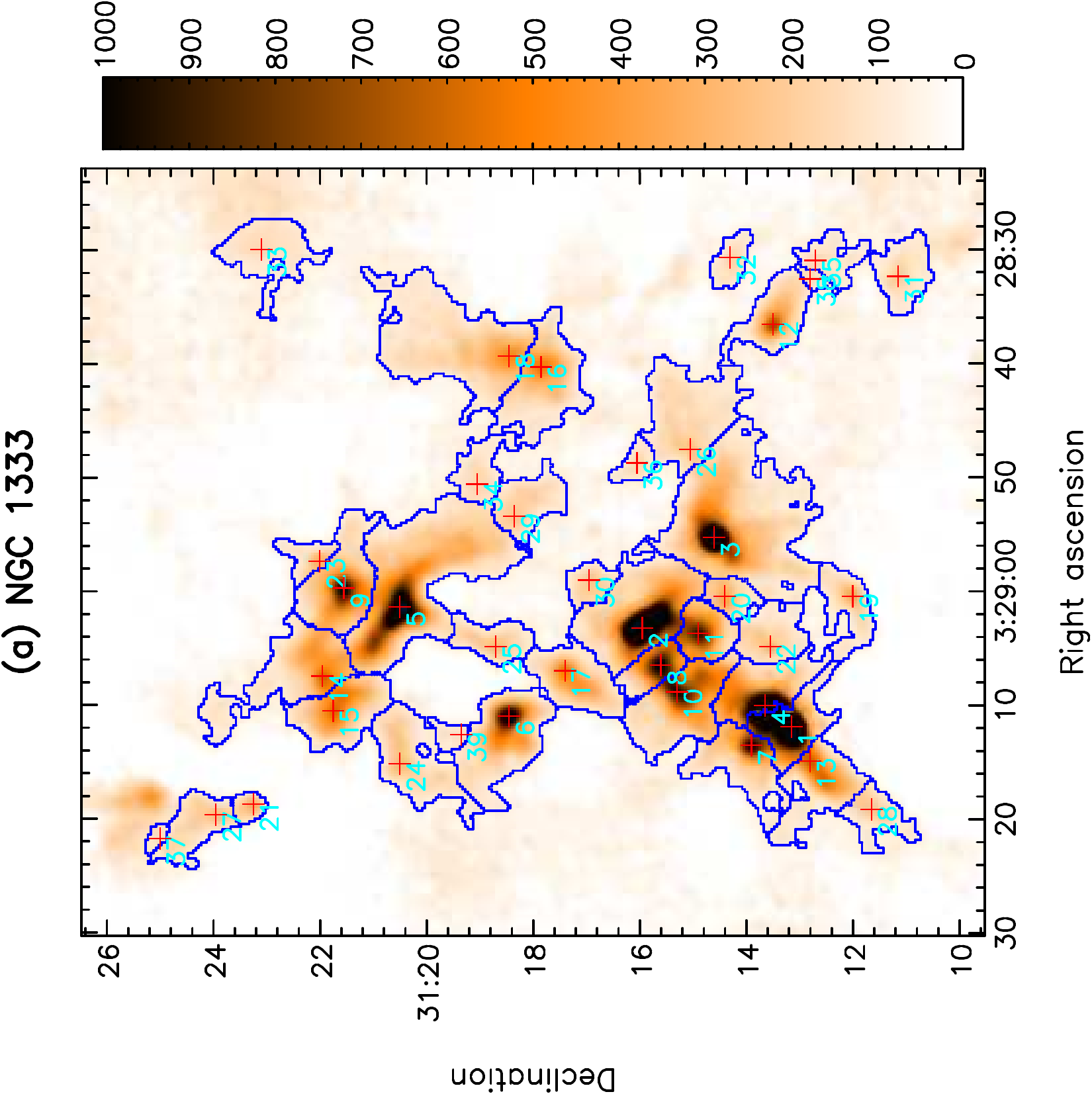}
\end{minipage}
\begin{minipage}{0.48\textwidth}
\includegraphics[height=\textwidth,angle=270]{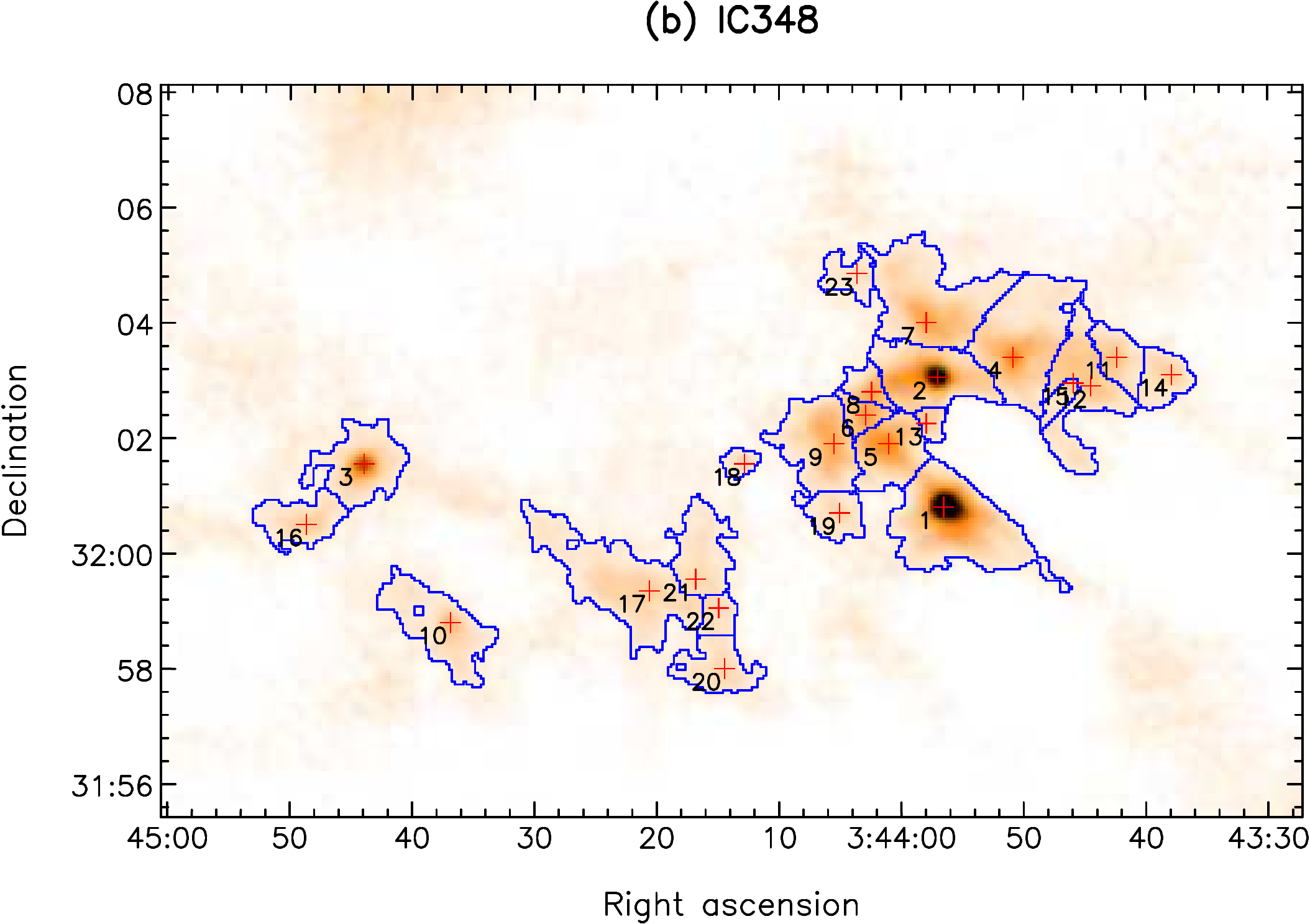}
\end{minipage}
\begin{minipage}{0.48\textwidth}
\includegraphics[height=\textwidth,angle=270]{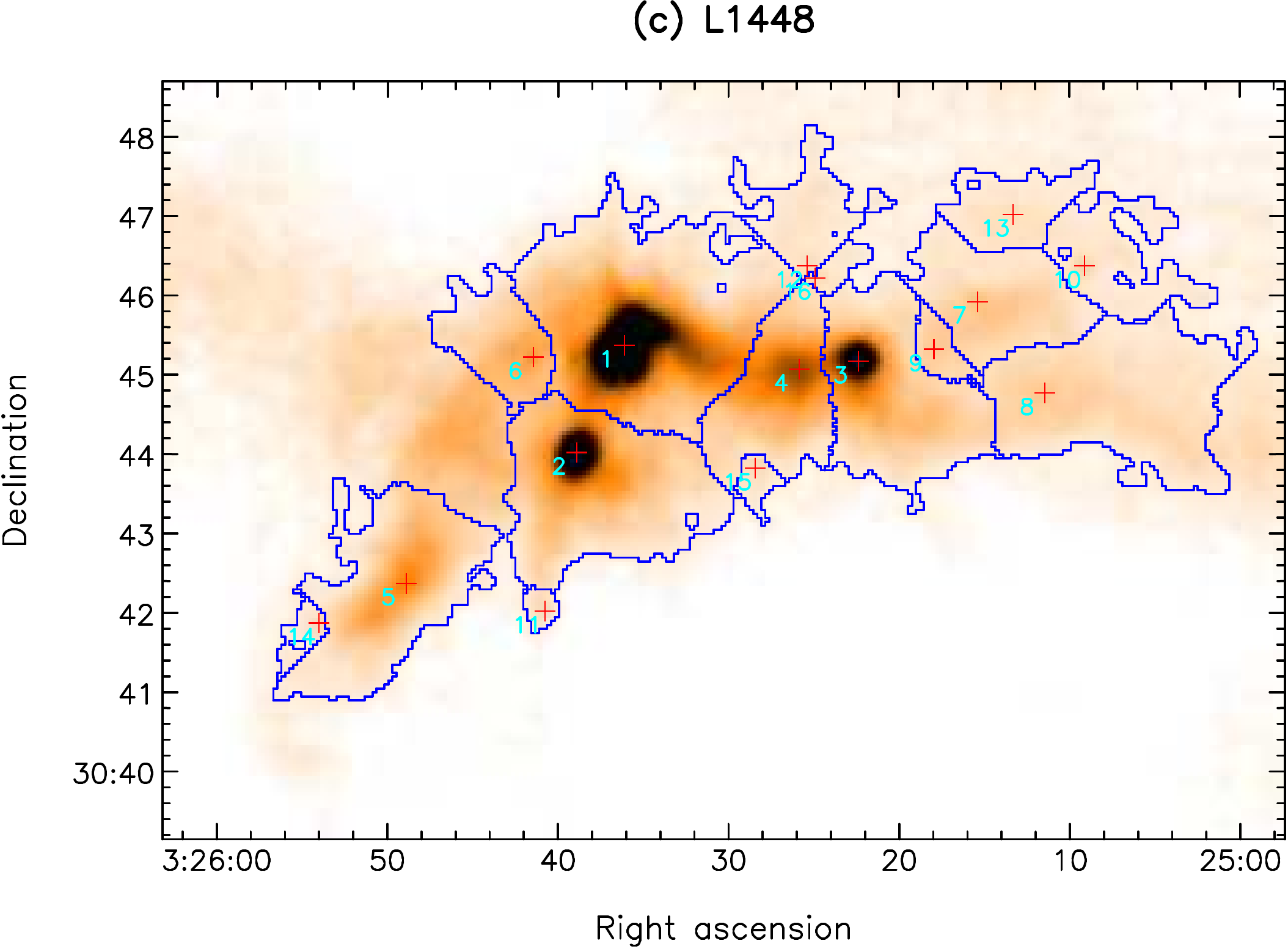}
\end{minipage}
\begin{minipage}{0.48\textwidth}
\includegraphics[height=\textwidth,angle=270]{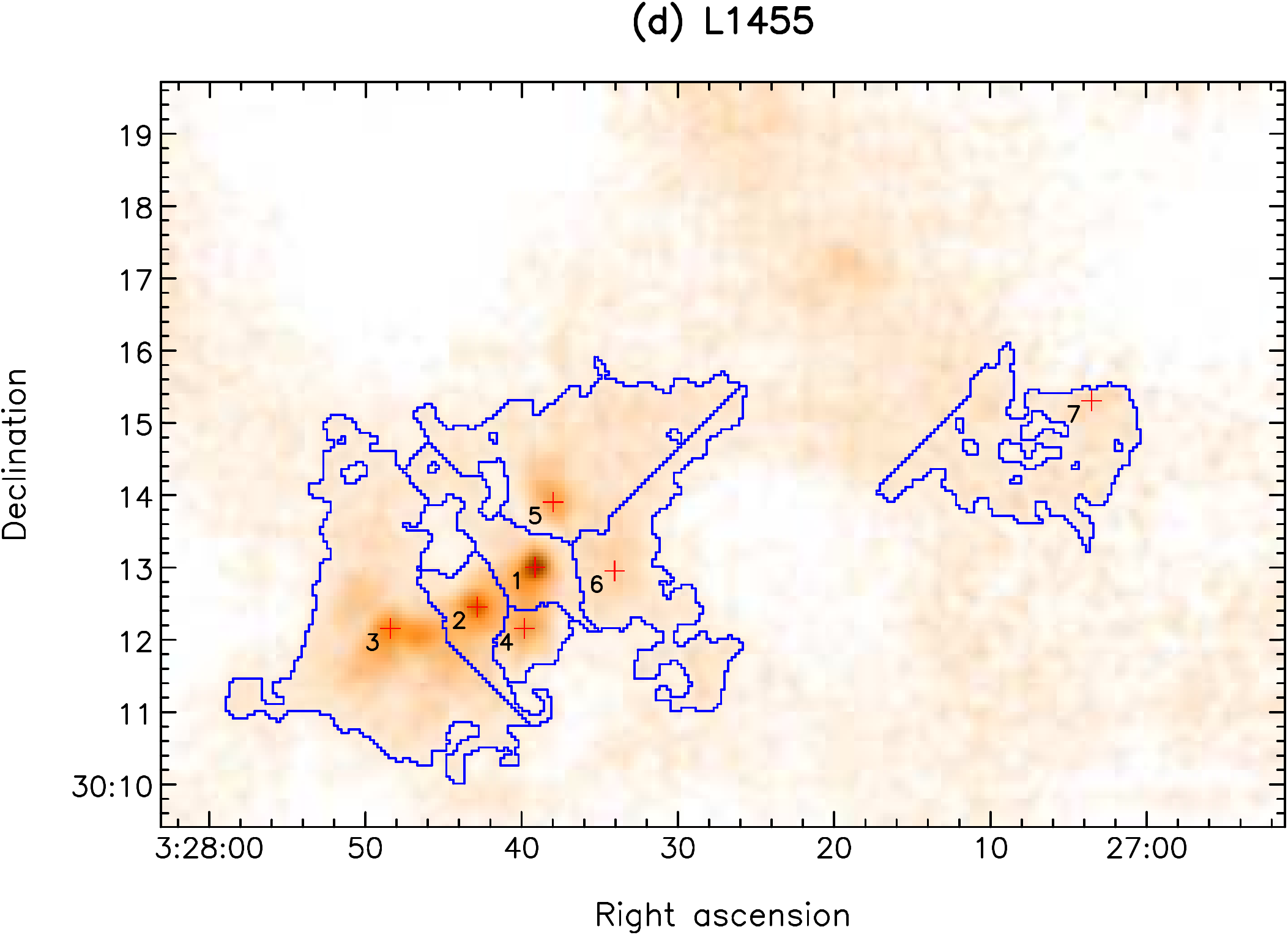}
\end{minipage}
\caption{Clumps found with \clfind\ in (a) NGC~1333, (b) IC348/HH211,
  (c) L1448 and (d) L1455. The regions selected are where we have
  complementary HARP data \citep{paper1}. The colour-scale is
  850\,\micron\ SCUBA flux density from 0 to
  1000\,\mjybeam. The contours show the \emph{boundaries} of the
  clumps and crosses mark the emission peaks, labelled with the clump number from the catalogue.}
\label{fig:clfind_clumps}
\end{center}
\end{figure*}

The most important parameters to tune \clfind's output are the value
of the lowest
contour level ($T_\rmn{low}$) and the contour level spacing ($\Delta
T$). $\Delta T$ of 2$\sigma_\mathrm{rms}$ is considered optimal for
rejecting clumps otherwise fitted to noise spikes
\citep{williams94}. $T_\mathrm{low}$ essentially sets the clump peak threshold ($T_\mathrm{low}+\Delta
T$). After some experimentation we settled on the
optimum parameters to find plausible clumps: $\Delta T$ and
$T_\rmn{low}$ both $2 \sigma_\rmn{rms}$, so clumps must have peaks
$\geq 4 \sigma_\rmn{rms}$. Additionally, the clumps had to be larger
than the beam size and contain at least 7
pixels. We find 85 clumps in this manner which are depicted across the
four regions in Fig.\ \ref{fig:clfind_clumps}. The full catalogue of
\clfind\ detections is listed in Tab.\ \ref{tab:clfind_detections}.

\subsubsection{\gclumps}

\begin{figure*}
\begin{center}
\begin{minipage}{0.48\textwidth}
\includegraphics[height=\textwidth,angle=270]{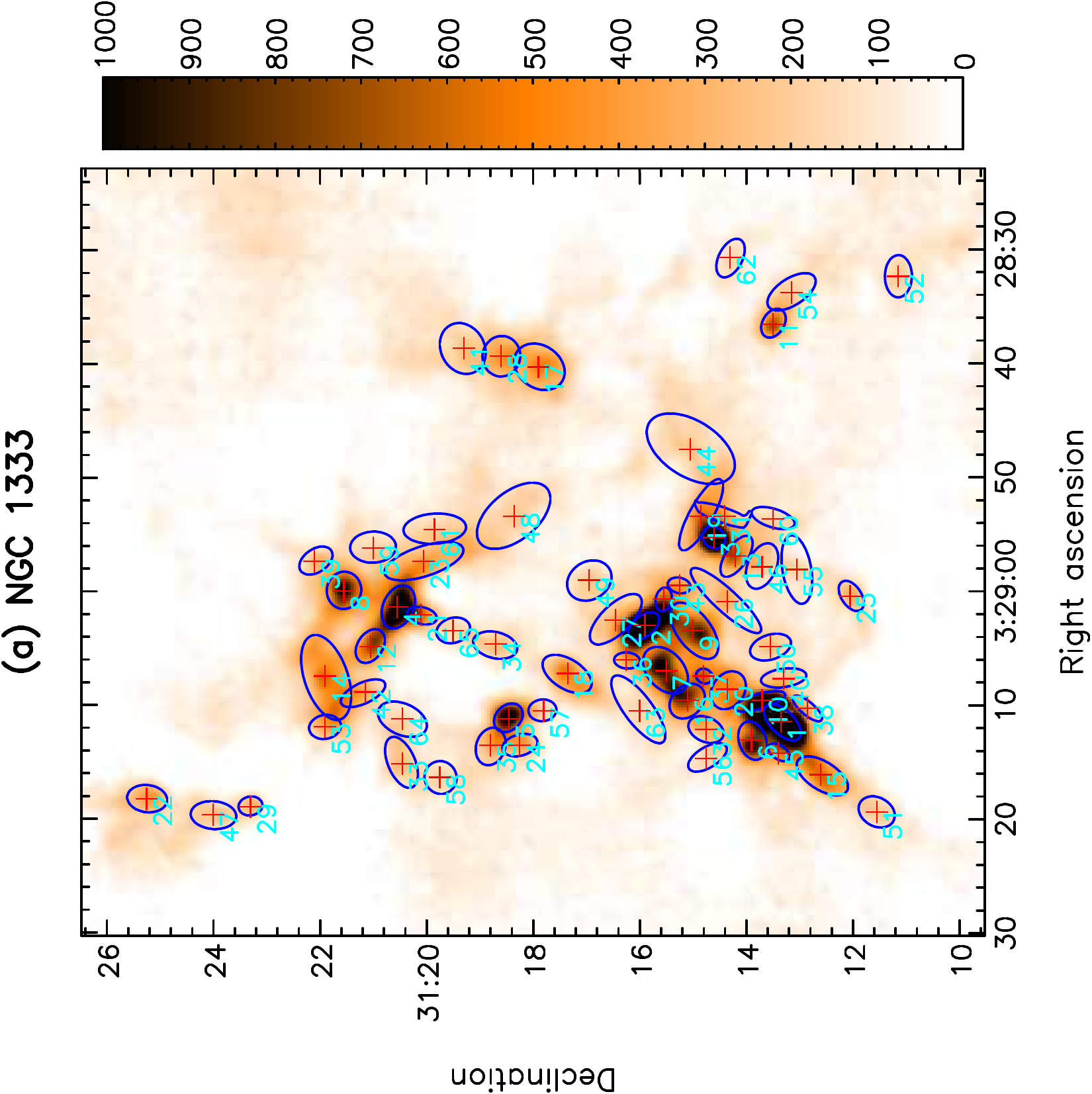}
\end{minipage}
\begin{minipage}{0.48\textwidth}
\includegraphics[height=\textwidth,angle=270]{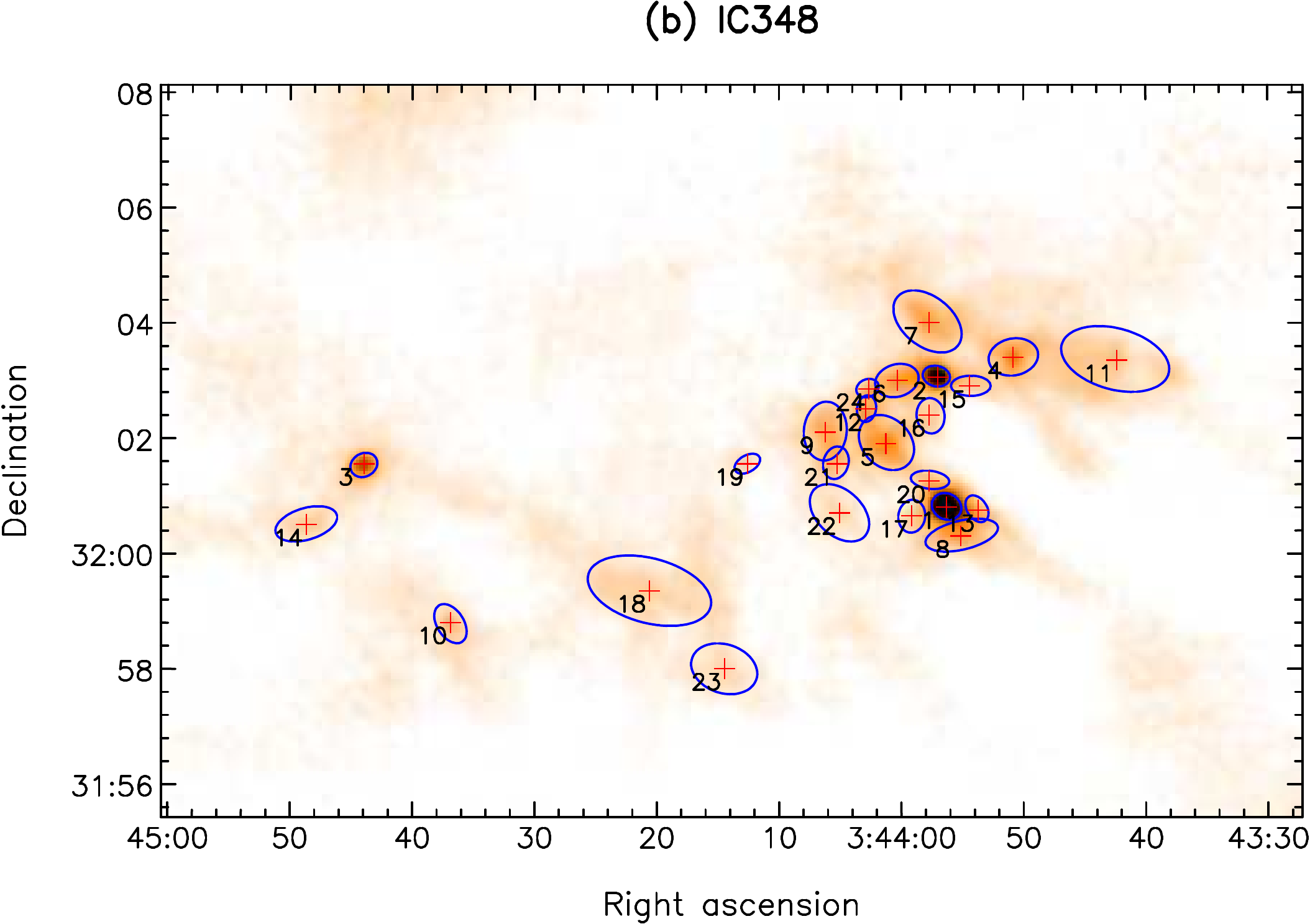}
\end{minipage}
\begin{minipage}{0.48\textwidth}
\includegraphics[height=\textwidth,angle=270]{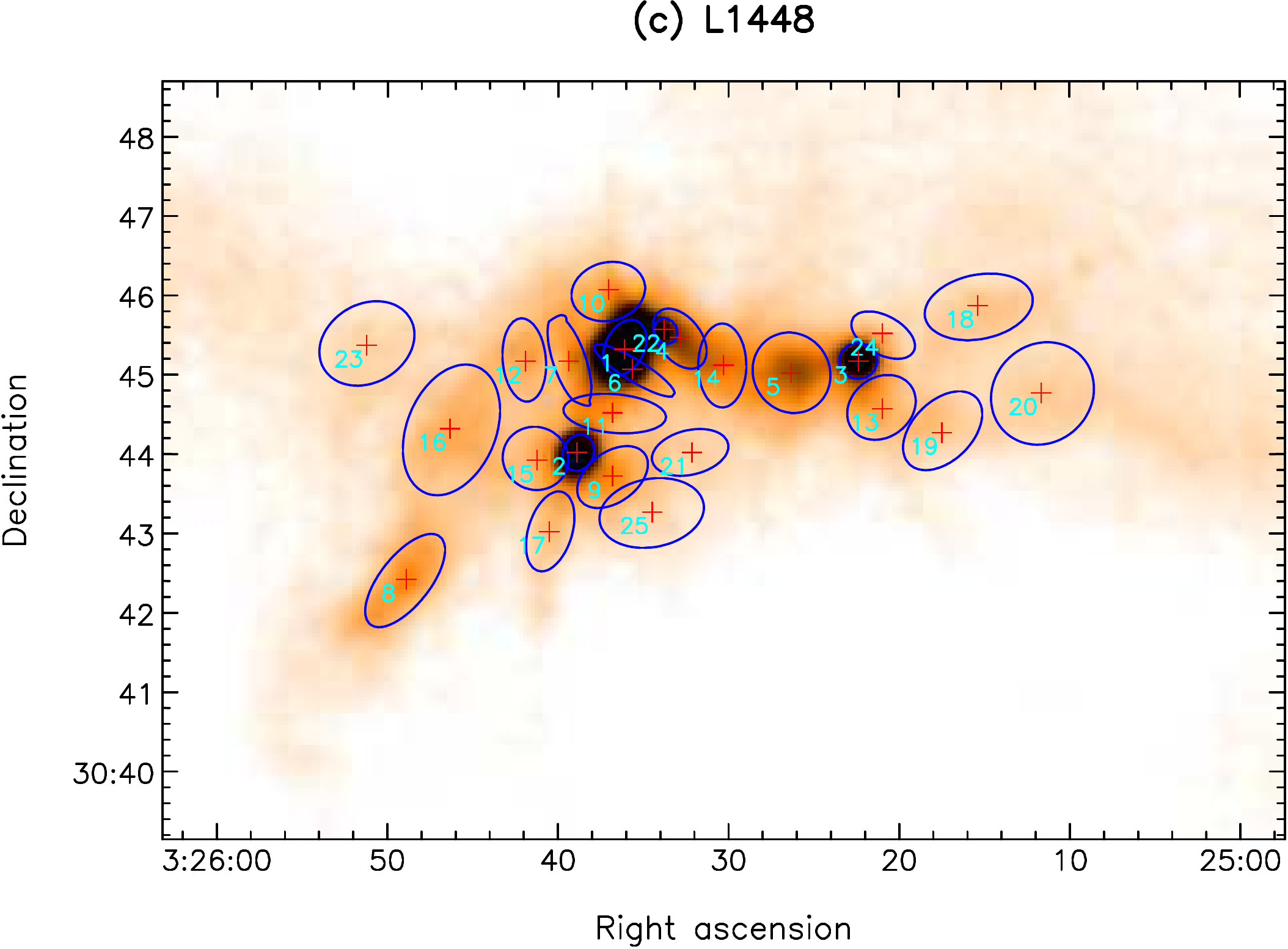}
\end{minipage}
\begin{minipage}{0.48\textwidth}
\includegraphics[height=\textwidth,angle=270]{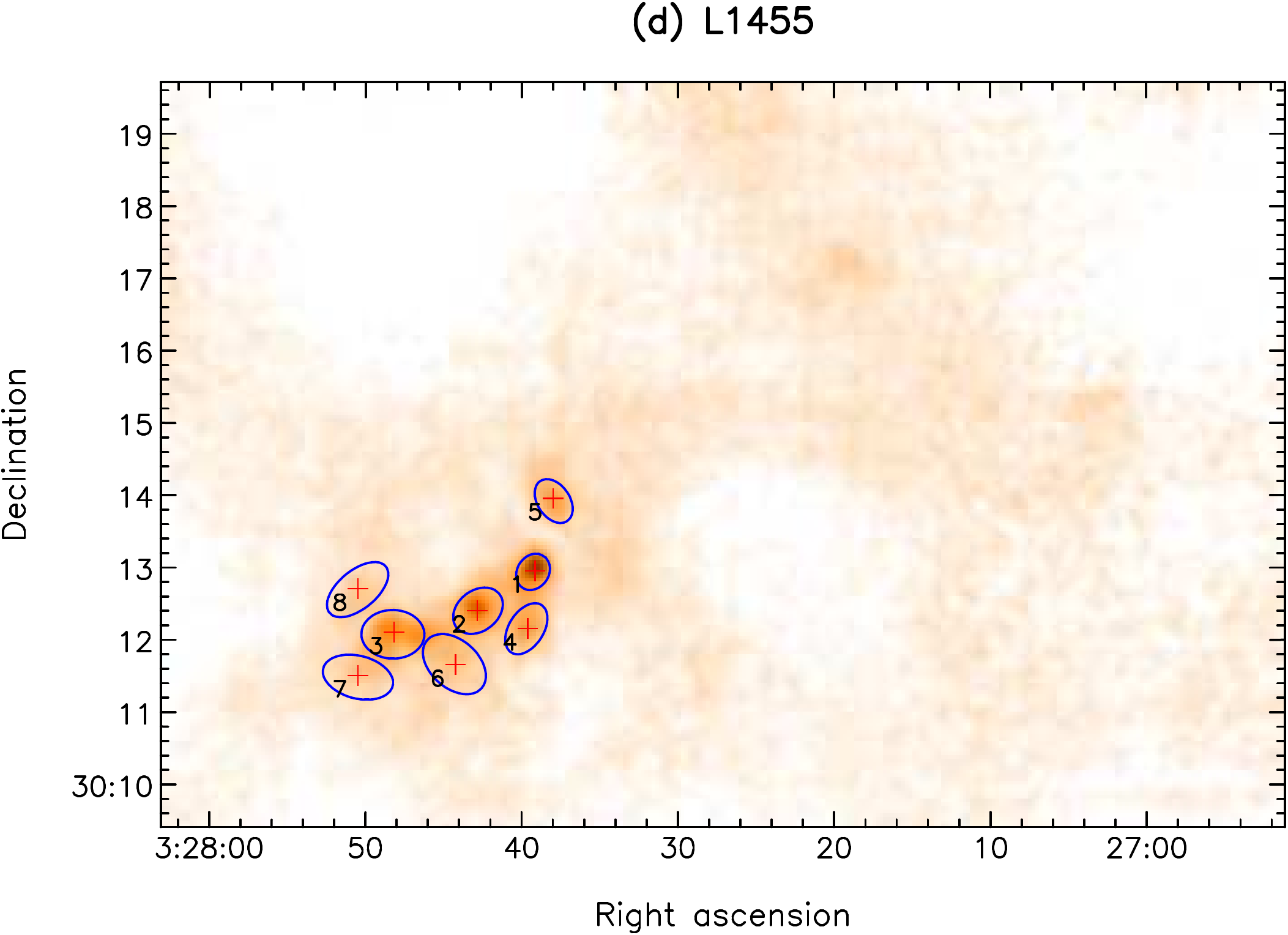}
\end{minipage}
\caption{Clumps found with \gclumps\ in (a) NGC~1333, (b) IC348/HH211,
  (c) L1448 and (d) L1455. The regions selected are where we have
  complementary HARP data \citep{paper1}. The colour-scale is
  850\,\micron\ SCUBA flux density from 0 to
  1000\,\mjybeam. The contours show the full-width half-maximum (FWHM) of the Gaussian model fit for each clump
  with crosses at the emission peak, labelled with the clump number from the catalogue.}
\label{fig:gauss_clumps}
\end{center}
\end{figure*}

For the \gclumps\ algorithm the most sensitive tuning
parameter was the threshold, the minimum value at which the routine would attempt a model fit ($4\sigma_\mathrm{rms}$). As
for \clfind, clumps with widths smaller than the beam size (14\,arcsec)
were rejected. In total, 122 clumps were located in the four regions as
shown in Fig.\ \ref{fig:gauss_clumps}. Tab.\
\ref{tab:gclumps_detections} provides the full the list of \gclumps\ identifications.

\subsubsection{Catalogue differences}

Although we identify clumps in the same data with the same algorithm,
\citetalias{hatchell07a} find fewer
clumps (the total number of clumps found in this work and
\citetalias{hatchell07a} is summarized in Tab.\ \ref{table:numberclumps_allalgs}). \citetalias{hatchell07a} use the {\sc idl} implementation of
\clfind\ on the SCUBA maps with an unsharp mask applied to remove
structure on spatial scales $>2$\,arcmin. The most plausible
explanation for this discrepancy is the different clump threshold,
although there are other differences. In our catalogue, we include all
\clfind\ detections found with $T_\mathrm{low}=2\sigma_\mathrm{rms}$
and $\Delta T=2\sigma_\mathrm{rms}$. This ensures the detections all have peaks $\geq
4\sigma_\mathrm{RMS}$. However, \citetalias{hatchell07a} only select
their \clfind\ detections (with $T_\mathrm{low} = 3\sigma_\mathrm{rms}$
and $\Delta T = \sigma_\mathrm{rms}$) that have peaks $\geq
5\sigma_\mathrm{rms}$. Additionally, they select less bright sources,
with peaks $\geq 3\sigma_\mathrm{rms}$ that also have an
identification by eye in their original paper \citep{hatchell05} or
are Bolocam sources \citep{enoch06}. 

\begin{table}
\caption{Number of clumps found for the best configuration of both
  \cupid\ algorithms alongside the number of sources in
  \citetalias{hatchell07a}.}
\begin{tabular}{lccccc}
\hline
Algorithm$\backslash$Region & L1455 & L1448 & IC348 & NGC~1333 & Total\\
\hline
\citetalias{hatchell07a} & 5 & 7 & 17 & 29 & 58\\
\clfind\    & 7 & 16& 23 & 39 & 85\\
\gclumps\    & 8 & 25& 24 & 65 & 122\\
\hline
\end{tabular}
\label{table:numberclumps_allalgs}
\end{table}

\subsection{Distinguishing starless and protostellar clumps} 
\label{sec:identify}

\begin{figure}
\begin{center}
\includegraphics[height=0.45\textheight]{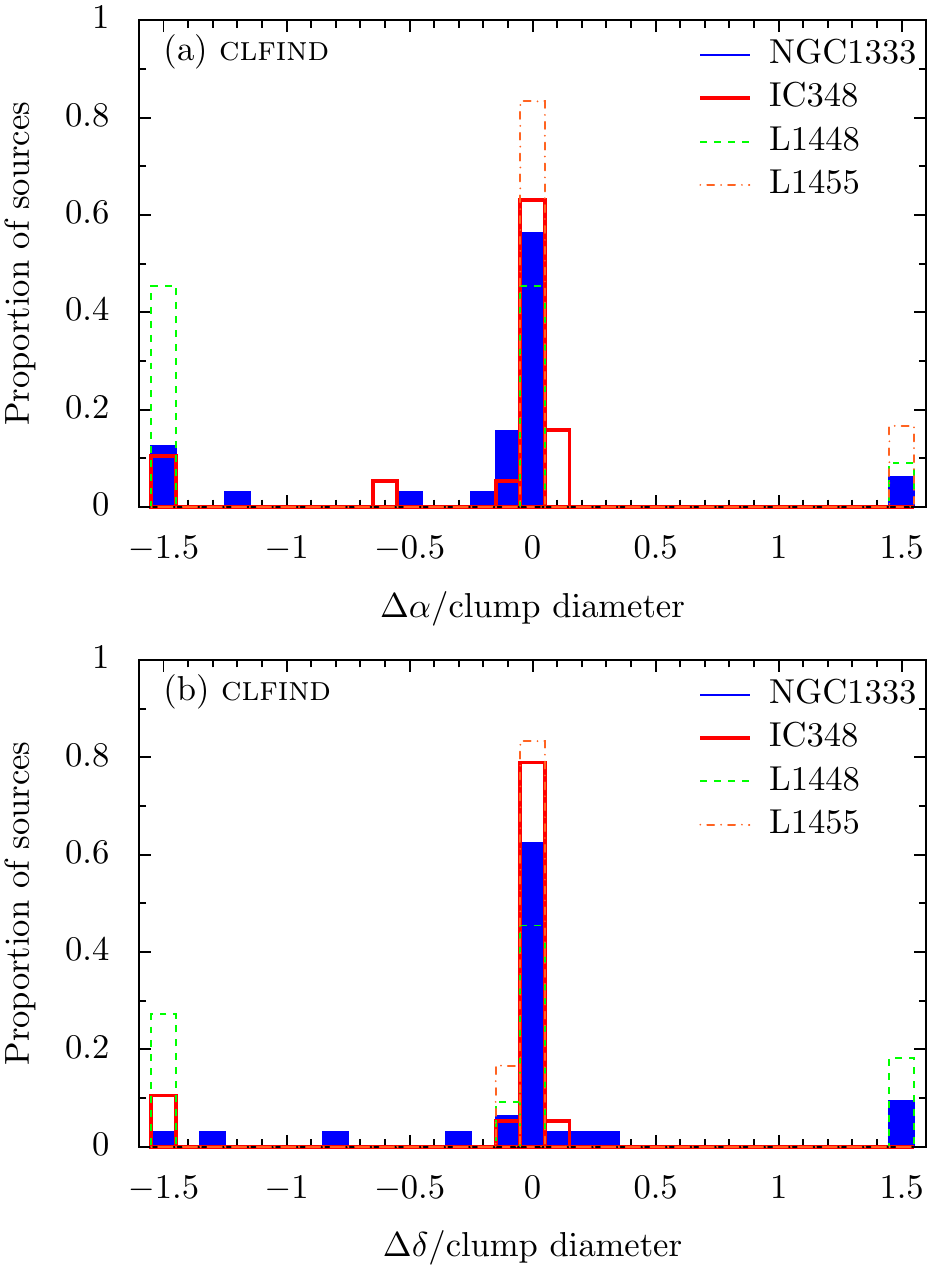}
\includegraphics[height=0.45\textheight]{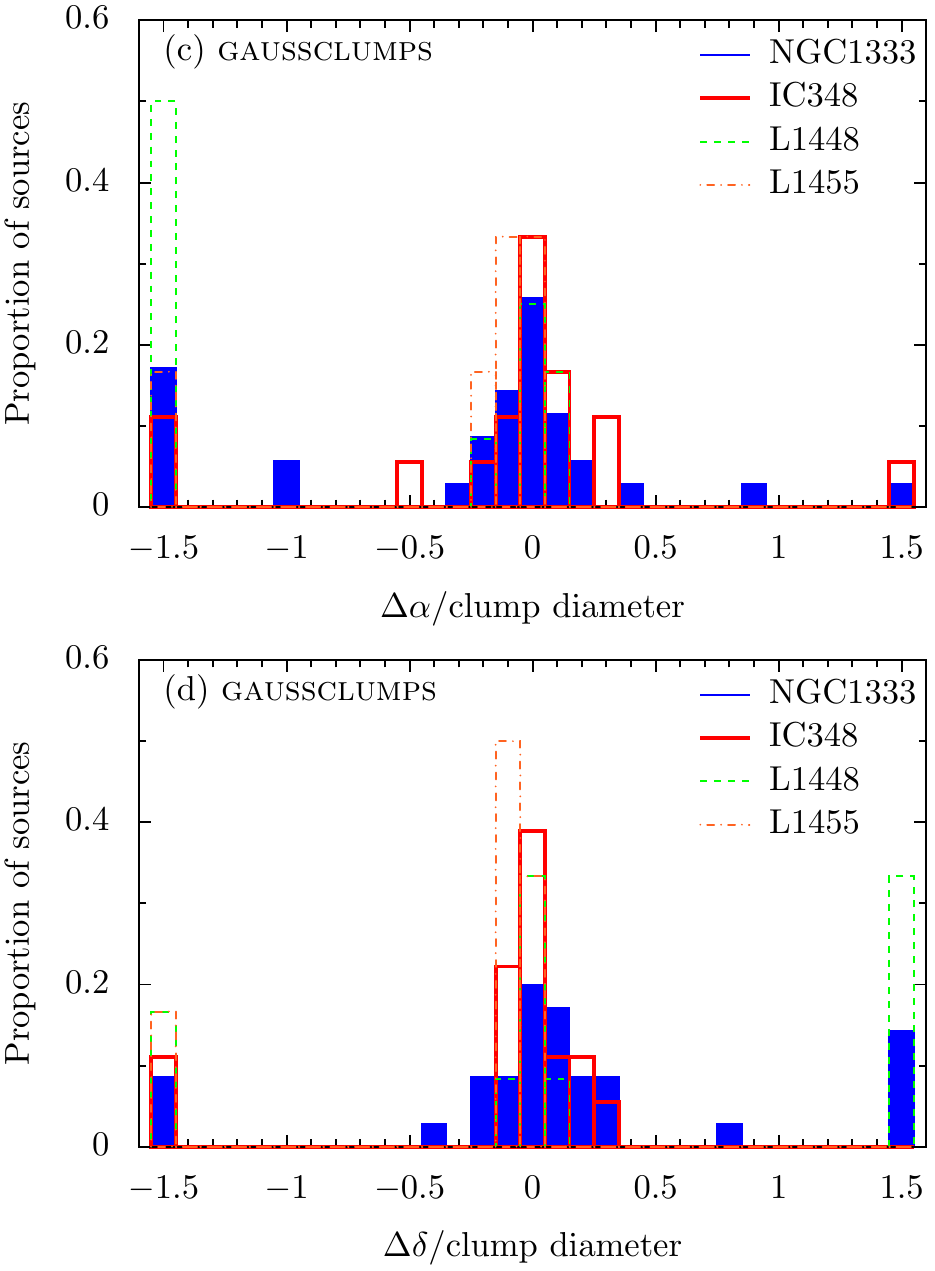}
\caption{Distribution of the positional
  offsets ($\Delta\alpha$,$\Delta\delta$) between each clump peak found with \clfind\ (top
  panels, (a) and (b)) or \gclumps\ (bottom panels, (c) and (d)) and the peak
  of its nearest \citetalias{hatchell07a} core by region. The distances are in units of the measured clump diameter. The
  alignment is almost perfect for \clfind\ and very good for
  \gclumps.}
\label{fig:sourcematching}
\end{center}
\end{figure}

To classify the clumps as starless or protostellar, we looked for an
associated source in the identifications of \citetalias{hatchell07a}. The positional offset from each of our clumps' peak to
the peak of its nearest \citetalias{hatchell07a} core in units of the
observed clump diameter is plotted
in Fig.\ \ref{fig:sourcematching} (cf. similar analysis of \citealp{enoch08}). The diameter along each axis is twice the clump
size, where the size is the standard deviation of the pixel coordinates
about the centroid, weighted by the pixel values. We associate clumps
with a core if they lie within one diameter. For \clfind,
53 matching clumps (62 per cent of the total) were located within one diameter of the
\citetalias{hatchell07a} source. All but two of these 53 are within half a
diameter. Indeed most of our clumps lie directly on top of the \citetalias{hatchell07a} sources so we can confidently classify them. The
\gclumps\ results are similar with 52 clumps (43
per cent) matching to within one diameter. All but four of these lie within
half a diameter. Three clumps
are paired with two sources, as their emission is sufficiently large
to encapsulate two \citetalias{hatchell07a} cores. The separation of the \citetalias{hatchell07a}
cores and Gaussian clumps is significantly larger than the near
perfect alignment of the \clfind\ clumps. However, the
associations are still unambiguous and far from a random distribution.  

\subsection{Dust mass estimates} 
\label{sec:dustmasses}

The mass of a clump, $M_{850}$, is proportional to its total
850\,\micron\ flux, $S_{850}$, provided the emission is optically thin and its only source is thermal
emission from the dust: \begin{equation}
  M_{\nu}=\frac{d^2S_\nu}{B_\nu(T_\mathrm{D})\kappa_\nu}\rmn{,} \end{equation} where
$\kappa_{\nu}$ is the dust opacity, $d$ the distance to the object and
$B_\nu$ the Planck function at $T_\mathrm{D}$, the dust
temperature. Here the conversion becomes: \begin{eqnarray} M_{850} & = &
  0.70~S_{850}\left[\exp\left(1.4\left(\frac{\mathrm{12\,K}}{T_\mathrm{D}}\right)\right)-1\right] \nonumber \\ & & \qquad \times \left(\frac{\kappa_{850}}{\mathrm{0.012\,cm^2\,g^{-1}}}\right)^{-1} \left(\frac{d}{\mathrm{250\,pc}}\right)^2\mathrm{M_\odot}\rmn{.}  \end{eqnarray} \label{eqn:dustmass}

As discussed in \citet{paper1}, we assume a constant distance of 250\,pc to
the entirety of the Perseus complex. Initially, we adopt
$T_\mathrm{D}=12$\,K everywhere. This temperature is consistent with
the typical kinetic temperatures ($T_\rmn{k}\sim11$\,K) derived by
\citet{rosolowsky08} from ammonia observations towards 193 dense core
candidates in Perseus and dust temperature measurements at 5\,arcmin
resolution of the whole of the cloud
\citep{schnee08b}. \citet{schnee08b} find Perseus uniformly cool
($T_\rmn{D}\sim$12--20\,K) with a median $T_\rmn{D}=15$\,K. When the
sources are separated into classes by age, we assume a slightly higher temperature for those with an internal source of
luminosity, i.e.\ protostars (15\,K), over starless cores
(10\,K). This is consistent with the average isothermal dust temperatures
found in protostellar radiative transfer models
(\citealp*{shirley02}; \citealp{young03}) and complementary studies
\citep{hatchell08,enoch08}. A final uncertainty in the mass comes from
$\kappa_\nu$. We adopt a value, $\kappa_\nu= 0.012$\,cm$^2$\,g$^{-1}$,
derived from the icy coagulated grain model (number 5) of
\citet{ossenkopf94}, which is consistent with
\citet{hatchell05}. \citet*{kirk06}, for example, prefer
0.02\,cm$^2$\,g$^{-1}$, which results in our calculated masses being smaller by a
factor of 1.7 for a source of identical flux density. 

\subsection{Completeness}

Completeness critically affects any CMD. At low masses, sensitivity limits cause a
reduction in the number of clumps found and this could be
confused with a turn-over in the mass
spectrum. Furthermore, this completeness limit is a function of the
clump mass, radius and profile. Massive, extended cores as well as all low-mass cores will be
difficult to detect. Assuming constant surface brightness, a clump the
size of the beam (radius 7\,arcsec) must have at least a mass of
$\sim$0.10\,\msun\ for detection. At the mean radius of
the \clfind\ clumps (21.4\,arcsec, 0.026\,pc) 0.95\,\msun\ is required, whilst an average sized (17.1\,arcsec, 0.021\,pc) \gclumps\
detection necessitates $M\geq 0.58$\,\msun. Uniform surface brightness is one of the simplest assumptions we could
make for the clump profile. In reality, dust condensations are
more centrally concentrated, with profiles well-fitted to Bonnor-Ebert
spheres (e.g.\ \citealp{johnstone00b}; \citealp*{kirkj05}). Higher concentrations
ensure lower clump masses for a given radius and peak flux, so the uniform surface
brightness estimate should be viewed as an upper limit.
 
\section{Population characteristics} \label{sec:properties}

In this section we explore the physical properties of the
populations, summarized in Tabs.\ \ref{tab:overallproperties_clfind}
and \ref{tab:overallproperties_gclumps}. We examine how a core evolves
over time by analysing the populations of different core ages, from
prestellar to Class I sources, looking for significant differences in
the distributions using Kolmogorov-Smirnov (K-S) tests. 

Overall, similar numbers of clumps are found for the two algorithms,
apart from in L1448
and NGC~1333, where \gclumps\ finds considerably more. \gclumps\ fits a strict Gaussian function, which is
less flexible in a crowded, high-flux-density area with many clumps
probably not following such a profile. Once one model has been
subtracted there will still be ample flux in the residuals to continue
to fit Gaussians, whereas a method such as \clfind\
may distribute this flux to a smaller number of high-mass clumps (e.g.\ \citealp{williams94,smith_clark08}). 

\subsection{Size} \label{sec:clump_sizes} 

\begin{figure}
\begin{center}
\includegraphics[width=0.47\textwidth]{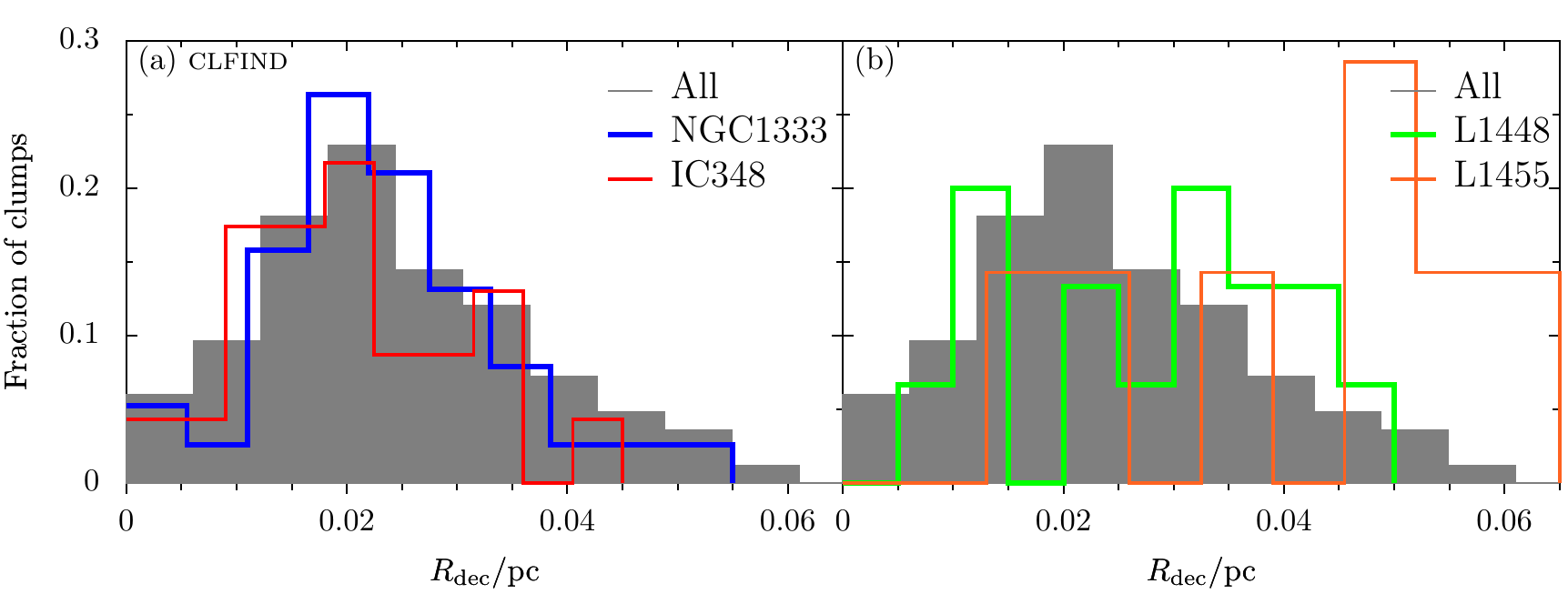}
\includegraphics[width=0.47\textwidth]{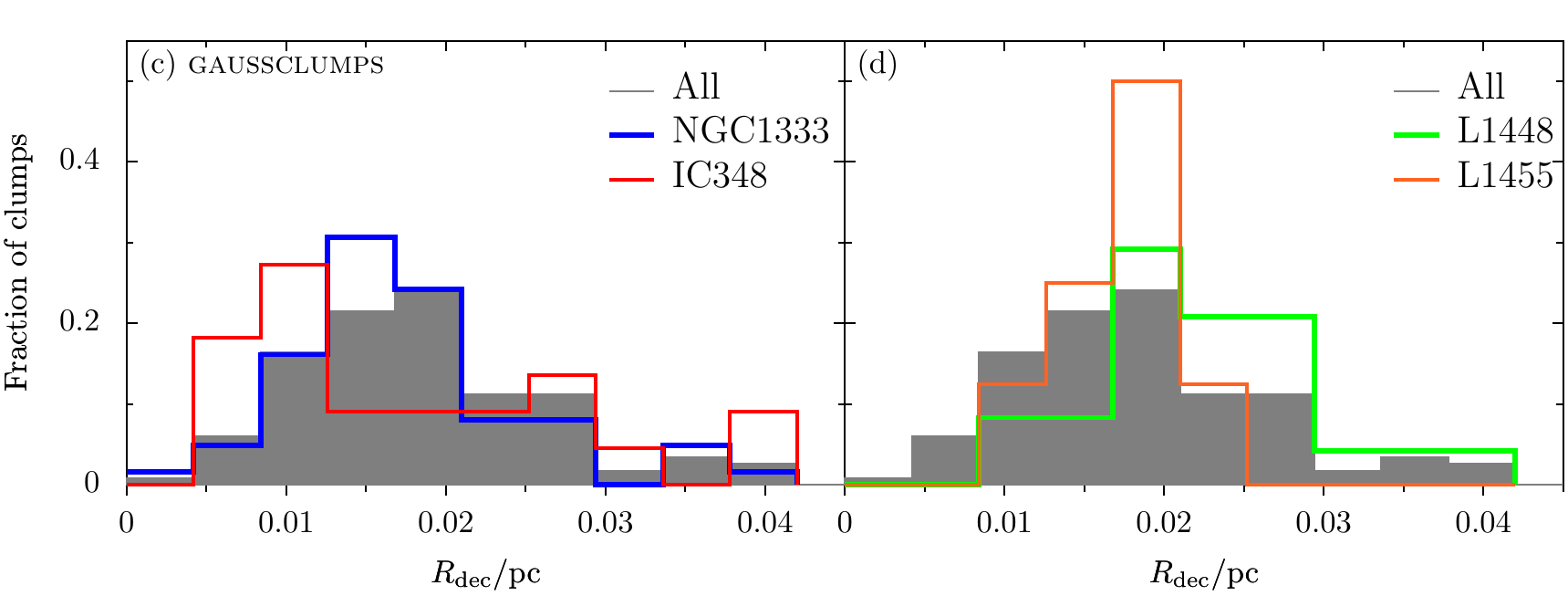}
\caption{Distribution of
  clump radii deconvolved with the beam size, $R_\mathrm{dec}$, broken
  down by region for the \clfind\ (top, (a) and (b)) and \gclumps\ (bottom,
  (c) and (d)) populations.}
\label{fig:deconvolved_sizesbyregion}
\end{center}
\end{figure}

\begin{figure}
\begin{center}
\includegraphics[width=0.47\textwidth]{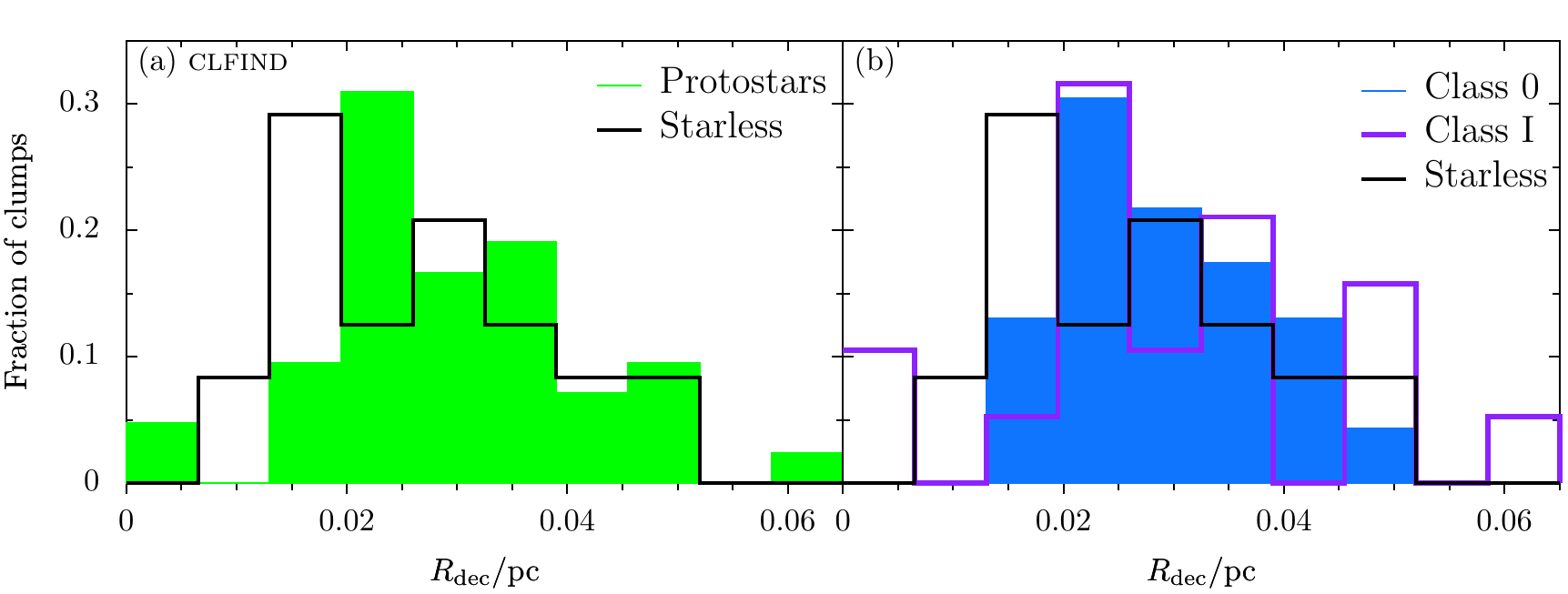}
\includegraphics[width=0.47\textwidth]{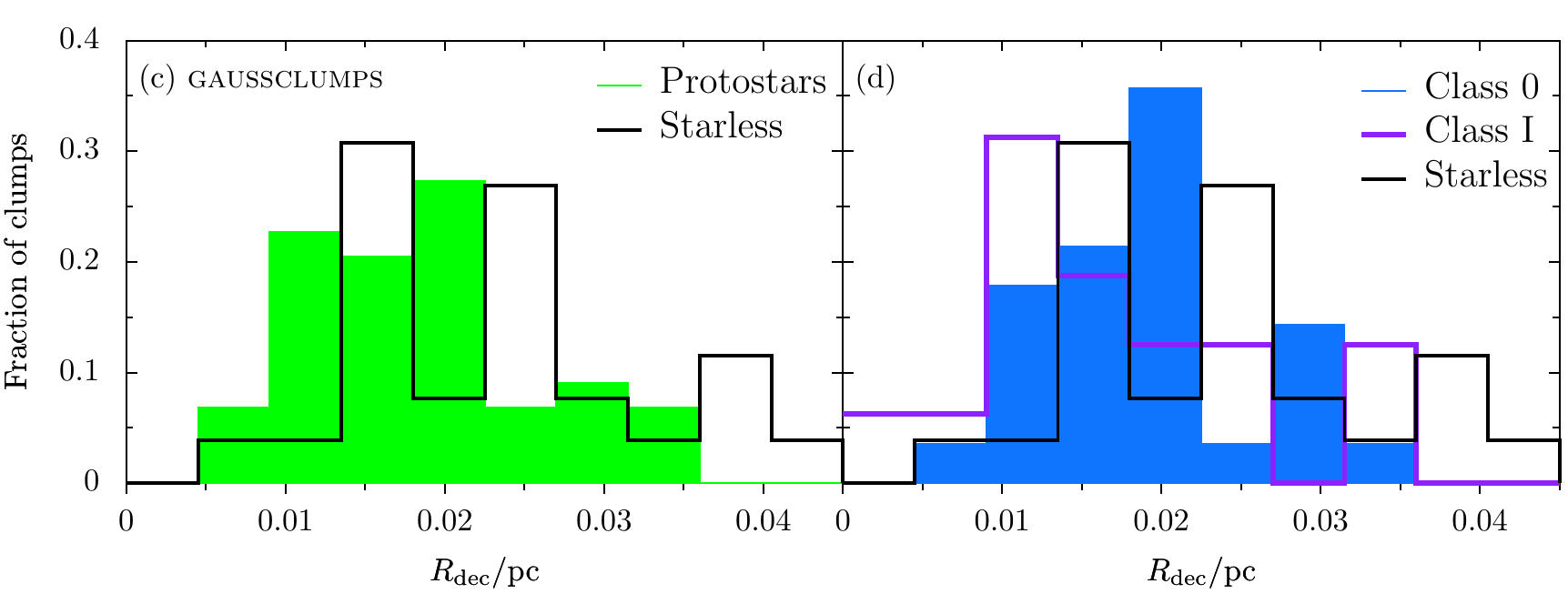}
\caption{Distribution of
  clump radii deconvolved with the beam size, $R_\mathrm{dec}$, for
  the various core classifications of \citetalias{hatchell07a} for the \clfind\ (top, (a) and (b)) and \gclumps\ (bottom,
  (c) and (d)) populations.}
\label{fig:deconvolved_sizesbysource}
\end{center}
\end{figure}

In Figs.\ \ref{fig:deconvolved_sizesbyregion} and \ref{fig:deconvolved_sizesbysource}, we plot histograms of
the clump deconvolved radius, $R_\mathrm{dec}$, formed as the geometric mean of
the clump major and minor axis sizes (see Section \ref{sec:identify}), each deconvolved
with the beam half-width half-maximum (HWHM) size ($R_\mathrm{beam}=$7\,arcsec or 0.0085\,pc):  
\begin{eqnarray} R_\mathrm{dec} & = & \left (
  R_\mathrm{dec,maj}.R_\mathrm{dec,min}\right )^{1/2} \nonumber \\ & =
    &\left (
    \sqrt{\mathrm{Size1}^2-{R_\mathrm{beam}}^2}.
    \sqrt{\mathrm{Size2}^2-{R_\mathrm{beam}}^2}\right )^{1/2} \end{eqnarray}
The \clfind\ population has a positively-skewed
distribution with mean $\langle R_\mathrm{dec}
\rangle=(0.0244\pm0.0014)$\,pc~\footnote{All the uncertainties listed
  throughout this paper are errors on the mean, formed from the
  sample standard deviations, $\sigma$, as $\sigma /\sqrt{N}$, where
  $N$ is the number in the sample.} (see Tab.\ \ref{tab:overallproperties_clfind}). There are some minor differences
region-to-region. IC348 and NGC~1333 have similar distributions to the
whole, the former skewed to slightly smaller sizes with $\langle R_\mathrm{dec} \rangle=(0.020\pm 0.002)$\,pc compared to
NGC~1333's $(0.0227\pm 0.0016)$\,pc. L1448 and L1455 have fewer clumps so
their distributions are uncertain, but they do appear flatter with means of $(0.027\pm0.003)$\,pc and
$(0.041\pm0.006)$\,pc respectively. K-S tests suggest the NGC~1333 population is significantly different from
those in L1448 and L1455 with probabilities of being drawn from the
same population of 7 and 2~per cent respectively. Similar results
are found for IC348 compared to L1448 and L1455 (5 and 2~per cent). 

\begin{table*}
\caption{Average clump properties for the \clfind\ population. All the errors listed ($\sigma$) are
  errors on the mean \textit{not} sample deviations.}
\begin{tabular}{lllllllll}
\hline
Population & $R_\rmn{dec}$\,$^\rmn{a}$ & $\sigma_R$ & AR\,$^\rmn{b}$ &
$\sigma_\rmn{AR}$ & $N(\rmn{H_2})$\,$^\rmn{c}$ & $\sigma_N$ & $M$ & $\sigma_M$ \\
& (pc) & (pc) & & & ($10^{23}$\,cm$^{-2}$) & ($10^{23}$\,cm$^{-2}$) &
(M$_\odot$) & (M$_\odot$)\\
\hline
All        & 0.0244 & 0.0014 & 1.50 & 0.07 & 1.40 & 0.18 & 6.4 & 0.8\\
NGC~1333   & 0.0227 & 0.0016 & 1.44 & 0.10 & 1.7  & 0.3  & 6.8 & 1.2\\
IC348/HH211& 0.020 & 0.002 & 1.61 & 0.17 & 0.93 & 0.18 & 3.6 & 0.7\\
L1448      & 0.027  & 0.003  & 1.50 & 0.15 & 1.5  & 0.5  & 9   & 3\\
L1455      & 0.041  & 0.006  & 1.42 & 0.10 & 0.93 & 0.16 & 8   & 2\\
Starless   & 0.027  & 0.003  & 1.46 & 0.09 & 1.00 & 0.15 & 6.0 & 1.4\\
Protostars\,$^\rmn{d}$ & 0.0294 & 0.0019 & 1.42 & 0.10 & 2.6  & 0.4  & 10.6 & 1.7\\
Class 0    & 0.0287 & 0.0017 & 1.37 & 0.06 & 3.4  & 0.5  & 12  & 3\\
Class I    & 0.030  & 0.003  & 1.47 & 0.19 & 0.9  & 0.2  & 7.6 & 1.8\\
\hline
\end{tabular}
\label{tab:overallproperties_clfind}
\begin{flushleft}
$^\rmn{a}$~Deconvolved radius, see Section~\ref{sec:clump_sizes}. \\
$^\rmn{b}$~Axis ratio, see Section~\ref{sec:shapes}. \\
$^\rmn{c}$~Peak beam-averaged column density, see Section~\ref{sec:850_columndensity}. \\
$^\rmn{d}$~The entire Class 0 and Class I protostellar population. \\
\end{flushleft}
\end{table*}

\begin{table*}
\caption{Average clump properties for the \gclumps\ population. All the errors listed ($\sigma$) are
  errors on the mean \textit{not} sample deviations.}
\begin{tabular}{lllllllll}
\hline
Population & $R_\rmn{dec}$\,$^\rmn{a}$ & $\sigma_R$ & AR\,$^\rmn{b}$ &
$\sigma_\rmn{AR}$ & $N(\rmn{H_2})$\,$^\rmn{c}$ & $\sigma_N$ & $M$ & $\sigma_M$ \\
& (pc) & (pc) & & & ($10^{23}$\,cm$^{-2}$) & ($10^{23}$\,cm$^{-2}$) &
(M$_\odot$) & (M$_\odot$)\\
\hline
All        & 0.0186 & 0.0007 & 1.62 & 0.07 & 1.11 & 0.12 & 3.8 & 0.3\\
NGC~1333   & 0.0176 & 0.0009 & 1.68 & 0.11 & 1.19 & 0.18 & 3.8 & 0.5\\
IC348/HH211& 0.018  & 0.002  & 1.69 & 0.15 & 0.85 & 0.16 & 3.0 & 0.5\\
L1448      & 0.0225 & 0.0013 & 1.50 & 0.10 & 1.3  & 0.3  & 5.1 & 0.8\\
L1455      & 0.017  & 0.002  & 1.25 & 0.07 & 0.74 & 0.11 & 2.3 & 0.3\\
Starless   & 0.023  & 0.002  & 1.48 & 0.10 & 0.90 & 0.09 & 5.2 & 0.7\\
Protostars\,$^\rmn{d}$ & 0.0182 & 0.0011 & 1.43 & 0.07 & 2.3  & 0.3  & 6.8 & 0.9\\
Class 0    & 0.0190 & 0.0012 & 1.40 & 0.10 & 2.7  & 0.3  & 7.6 & 1.1\\
Class I    & 0.017  & 0.002  & 1.49 & 0.09 & 0.72 & 0.15 & 5.4 & 1.3\\
\hline
\end{tabular}
\label{tab:overallproperties_gclumps}
\begin{flushleft}
$^\rmn{a}$~Deconvolved radius, see Section~\ref{sec:clump_sizes}. \\
$^\rmn{b}$~Axis ratio, see Section~\ref{sec:shapes}. \\
$^\rmn{c}$~Peak beam-averaged column density, see Section~\ref{sec:850_columndensity}. \\
$^\rmn{d}$~The entire Class 0 and Class I protostellar population. \\
\end{flushleft}
\end{table*}

The trends in the \gclumps\ data are similar (see Tab.\ \ref{tab:overallproperties_gclumps}). The sizes
are smaller than those for \clfind\ by $\sim$0.005\,pc
with an overall mean, $\langle R_\mathrm{dec} \rangle=(0.0186\pm
0.0007)$\,pc. The IC348 and NGC~1333 clumps have similar means,
$(0.018\pm 0.002)$ and $(0.0176\pm 0.0009)$\,pc respectively. The L1448 population is slightly negatively
skewed with the highest average, $\langle R_\mathrm{dec} \rangle=(0.0225\pm
0.0013)$\,pc. K-S tests yield only 1, 3 and 4 per cent likelihoods that
the L1448 population is drawn from the same one as NGC~1333, IC348 and L1455
respectively. Clumps in L1455 have changed greatly from
their \clfind\ distribution; they now have the smallest average, $\langle R_\mathrm{dec} \rangle=(0.017\pm
0.002)$\,pc. 

For \clfind\, the starless clumps are marginally smaller than the protostellar clumps, $\langle R_\mathrm{dec} \rangle=(0.027\pm
0.003)$ compared to $(0.0294\pm 0.0019)$\,pc. The Class 0 and Class I
sources have nearly identical means ($(0.0287\pm 0.0017)$ and $(0.030\pm
0.003)$\,pc) and similar distributions with the Class 0s
narrower. Intriguingly, with \gclumps\ the opposite trend is
clear: on average starless cores are larger than Class 0 sources, which in
turn are larger than Class I sources ($\langle R_\mathrm{dec} \rangle=(0.023\pm 0.002)$,  $(0.0190\pm
0.0012)$ and $(0.017\pm 0.002)$\,pc respectively). 

What do these results reveal about star-forming cores? In each region we might expect the clump sizes to vary according
to the nature of the underlying population (i.e.\ protostellar or
starless) and degree of clustering. Tab.\
\ref{table:region_diagnostics} quantifies
these effects, through the mean density of sources and the percentage of
protostars. For \clfind\ the small clumps are in NGC~1333 and
IC348 with larger clumps in L1455 and L1448. NGC~1333 is the most clustered environment so we would expect
smaller clumps, since any measured size is limited by the
distance to the nearest source. IC348 is an exception (low source density with small clumps), but its cores
are significantly younger with many more designated starless that
dominate the overall size statistics (\clfind\ finds
starless clumps are smaller than protostars). \gclumps\ should be better at
separating blended sources and indeed the clustering in the regions
seems to have little bearing on the calculated sizes. The general trend in the
population is for starless cores to be larger than protostars. This is
the result reported by the Bolocam team \citep{enoch06,enoch08}
in their larger Perseus survey at 1.1\,mm with a similar
Gaussian fitting technique. Interestingly, they found the two types had similar
sizes in Serpens and Ophiuchus. 

Clearly the type of clump analysis makes a big difference to the
interpretation, a K-S test yields
only a 1~per cent chance that the two distributions arise from the
same population. In general \clfind\ finds larger clumps than
the tight spread of \gclumps\ radii. \clfind\ produces arbitrary shaped
clumps, rather than fitting a strict Gaussian profile. Therefore we might expect
more emission at larger distances thus increasing the radii
calculated. 

\begin{table*}
\caption{Clump densities and protostellar fractions. The protostellar fraction is
  the percentage of Class 0 and we sources out of the total
  clumps associated with \citetalias{hatchell07a} cores.}
\begin{tabular}{lcccccccccccc}
\hline
Region & Area & \multicolumn{2}{c}{Number of clumps} &
\multicolumn{2}{c}{Mean clump density} &
\multicolumn{2}{c}{Percentage of protostars} \\
& /arcmin$^2$ & & & \multicolumn{2}{c}{(arcmin$^{-2}$) } &
\multicolumn{2}{c}{in identified sources} \\
 & & \clfind\ &
\gclumps\ & \clfind\ &
\gclumps\ & \clfind\ &
\gclumps\ \\ 
\hline
NGC~1333 & 190.0 & 39 & 65 & 0.21 & 0.34&
76.9 & 72.4 \\
IC348 & 198.6 & 23 & 24 & 0.12 & 0.12 & 29.4 & 31.3 \\
L1448 & 111.1 & 16 & 25 & 0.14 & 0.23 & 80.0 & 71.4 \\
L1455 & 127.1 & 7 & 8 & 0.06 & 0.06 & 80.0 & 80.0 \\
\hline
\end{tabular}
\label{table:region_diagnostics}
\end{table*}

\subsection{Shape} \label{sec:shapes}

The shapes of star-forming cores provide clues to their formation. Models with cores forming from turbulent flows predict
random triaxial shapes with a slight preference for prolateness
\citep{gammie03,li04}. If strong magnetic fields are present
then cores are expected to be oblate
\citep{basu04,ciolek06}, which could also be caused by strong
rotational motion. Determining the exact 3-dimensional shape of a
core from observations is impossible, instead statistical techniques have
to be applied. Early axisymmetrical work indicated a preference for prolate cores
(e.g.\ \citealp{myers91}) but later studies consistently favour triaxial, preferentially
oblate shapes (e.g.\ \citealp*{jones01,goodwin02}; \citealp{tassis07,tassis09}). 

\begin{figure}
\begin{center}
\includegraphics[width=0.47\textwidth]{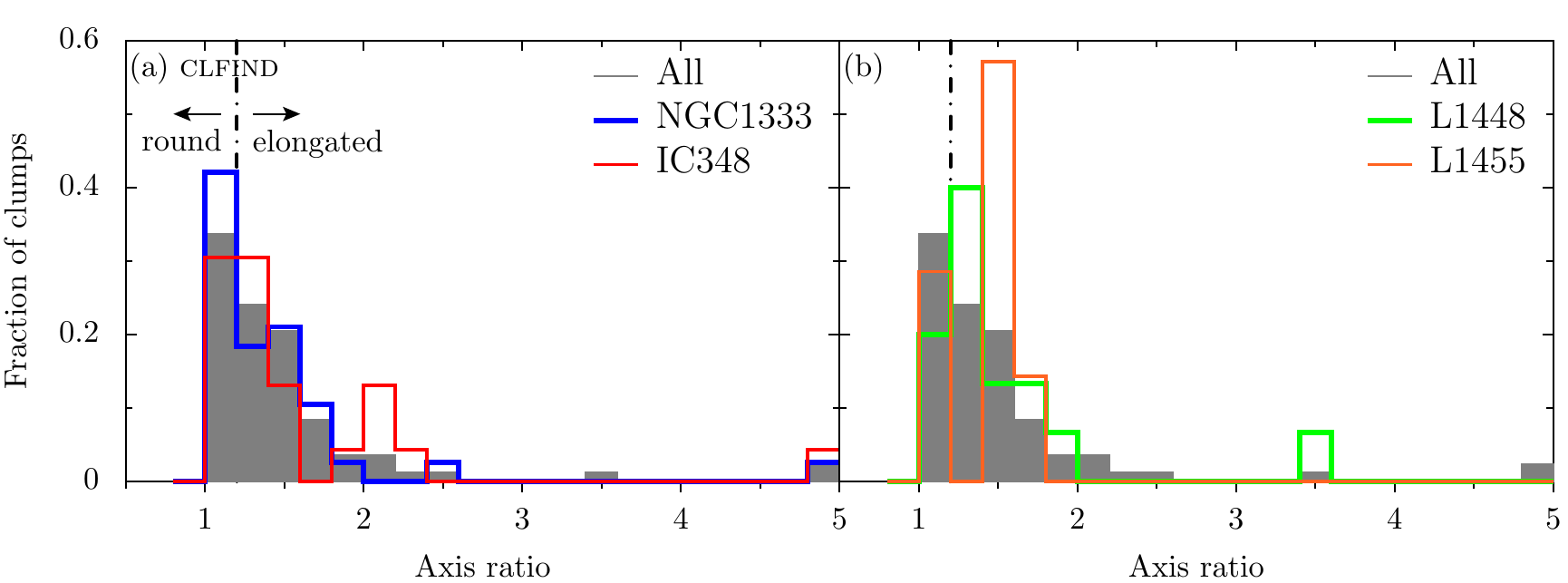}
\includegraphics[width=0.47\textwidth]{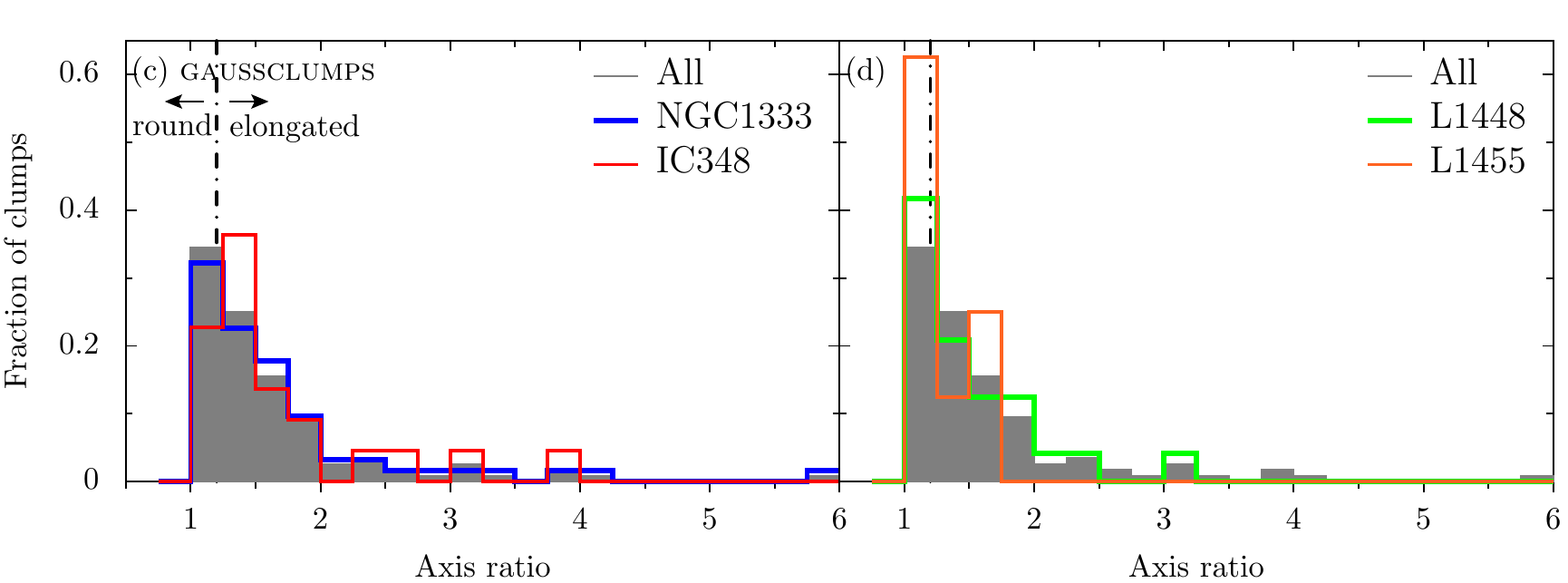}
\caption{Distribution of clump axis ratios by broken down by region for the \clfind\
  (top, (a) and (b)) and \gclumps\ (bottom, (c) and (d))
  populations. The dot-dashed vertical line at a ratio of 1.2
  separates clumps \citet{enoch06} consider round from those that are elongated.}
\label{fig:axisratios_byregion}
\end{center}
\end{figure}

\begin{figure}
\begin{center}
\includegraphics[width=0.47\textwidth]{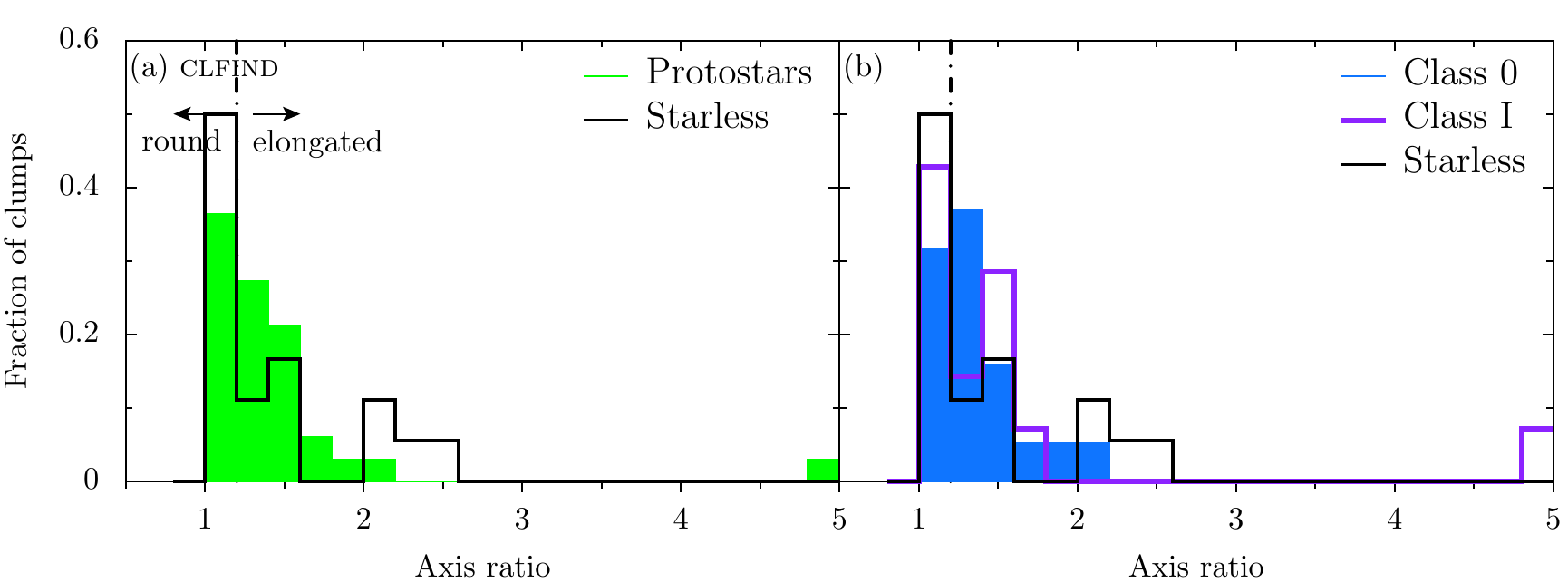}
\includegraphics[width=0.47\textwidth]{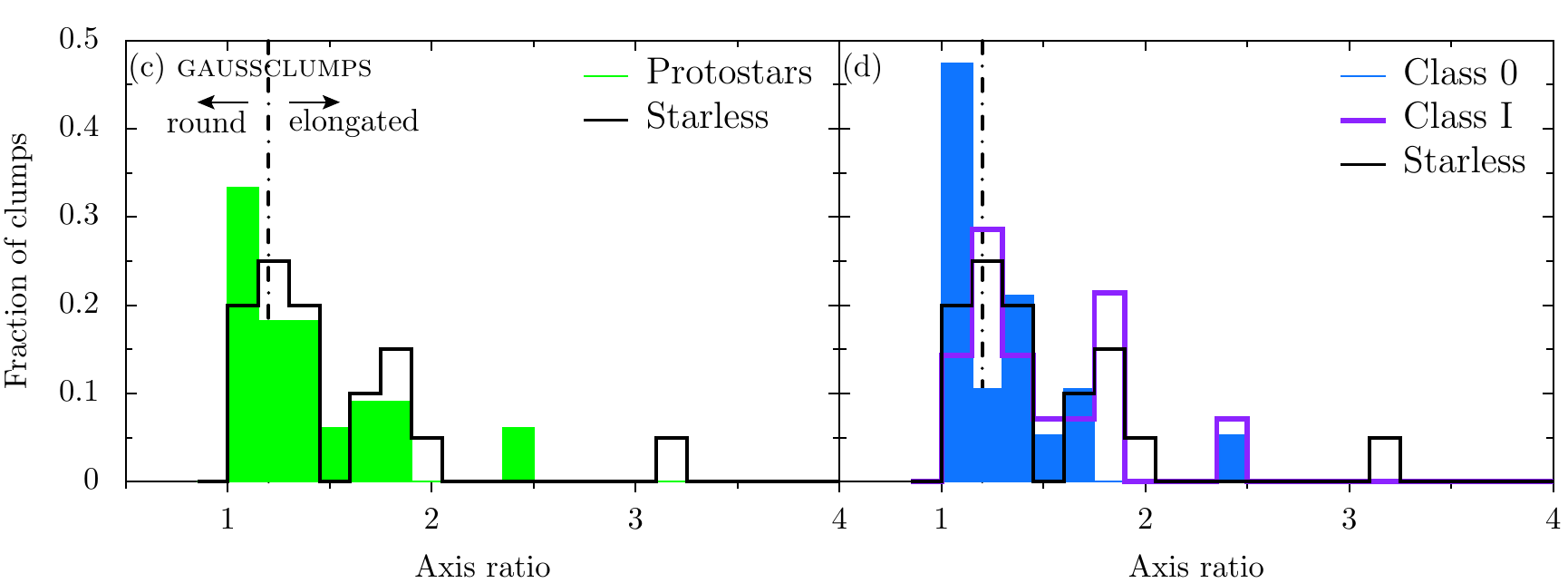}
\caption{Distribution of axis ratios by the source classifications of
  \citetalias{hatchell07a} for the \clfind\ (top, (a) and (b)) and \gclumps\
  (bottom, (c) and (d)) populations. The dot-dashed vertical line at a
  ratio of 1.2 separates clumps \citet{enoch06} consider round from
  those that are elongated.}
\label{fig:axisratios_bytype}
\end{center}
\end{figure}

A complete statistical analysis is beyond the scope of this paper,
instead we examine the clump ellipticity
in 2-dimensions, quantified via the axis
ratio. The axis ratio is the 
ratio of the major to minor clump radii each deconvolved with the beam,
$R_\mathrm{dec,maj}/R_\mathrm{dec,min}$. We plot the distribution of axis ratios in Figs.\
\ref{fig:axisratios_byregion} and \ref{fig:axisratios_bytype}, whilst
summarizing the data in Tabs.\ \ref{tab:overallproperties_clfind} and
\ref{tab:overallproperties_gclumps}. \citet{enoch06} performed
Monte Carlo simulations of a population of Gaussian sources to see the
effects of noise and mapping technique on their Bolocam data in
Perseus, which have a similar signal-to-noise ratio as ours. They conclude that a round underlying source
can be distorted to reach a maximum axis ratio of 1.2.

Region-to-region the distributions are similar, with
K-S tests unable to reject the hypothesis of identical populations to
any high level of significance. Most of the distributions have a positive skew
with a large proportion of clumps having ratios close to unity. On average, the clumps are slightly elliptical with ratios of
$1.5\pm 0.07$ for \clfind\ and $1.62\pm 0.07$ for
\gclumps. Starless and protostellar clumps have little
difference in their degree of elongation with ratios of $1.42\pm 0.10$
and $1.46\pm 0.09$ for \clfind\ plus $1.48\pm 0.10$ and $1.43\pm
0.07$ for \gclumps\ respectively. Class I sources appear
marginally more elongated than Class 0, though the distribution is affected by 
a few extremely elliptical sources: average ratios of $1.37\pm
0.06$ and $1.47\pm 0.19$ for Class 0 and I objects with \clfind\
alongside $1.40\pm 0.10$ and $1.49\pm 0.09$ for the same with
\gclumps. 

\citet{zeldovich70} and many others since have shown that a
dynamically collapsing triaxial ellipsoid will collapse fastest along
its shortest axis. Thus as collapse ensues axis
ratios increase. This is more or less the picture we see
here, with cores becoming slightly more elliptical as they
evolve. Surprisingly, \citet{goodwin02} found the opposite
trend with starless cores more elliptical than protostellar ones in
their more sophisticated statistical treatment of the deprojected shape of
\citet*{jijina99}'s N$_2$H$^+$ cores. This they tentatively propose as
evidence of magnetic fields complicating simple gravitational collapse
or of an outer protostellar \emph{envelope} that is not collapsing at all. 

\subsection{Column density} \label{sec:850_columndensity}

\begin{figure}
\begin{center}
\includegraphics[width=0.47\textwidth]{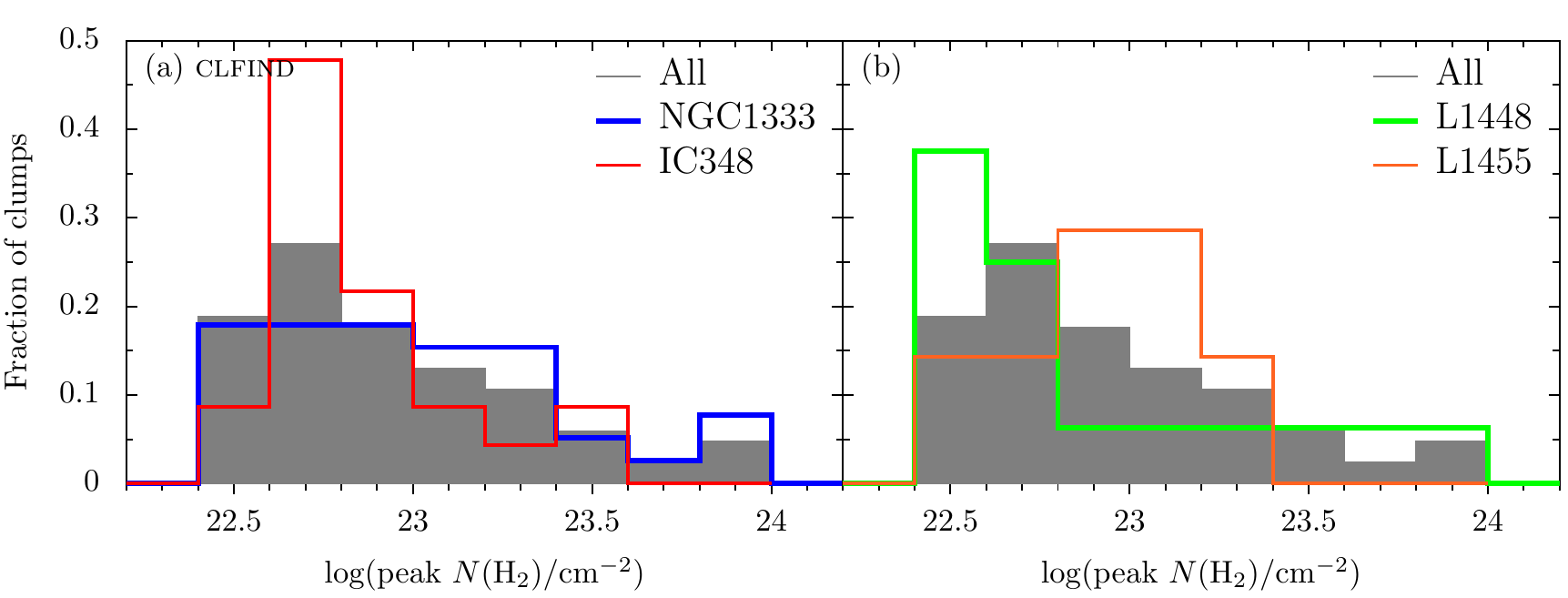}
\includegraphics[width=0.47\textwidth]{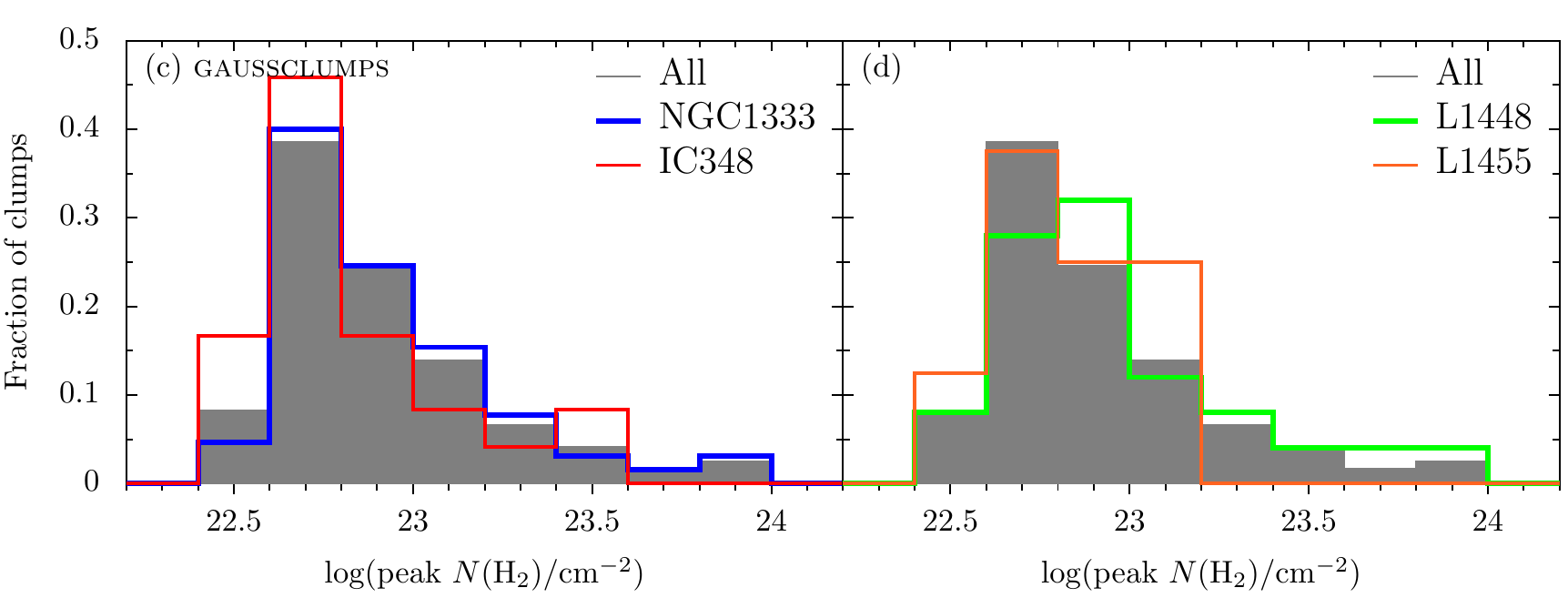}
\caption{Distribution of clump peak beam-averaged column densities for the different regions
  and overall for the \clfind\ (top, (a) and (b)) and \gclumps\
  (bottom, (c) and (d)) populations. The densities were
  calculated assuming a constant dust temperature of 12\,K. All regions exhibit similar behaviour apart from L1455
  which has the fewest clumps.}
\label{fig:column_byregion}
\end{center}
\end{figure}

\begin{figure}
\begin{center}
\includegraphics[width=0.47\textwidth]{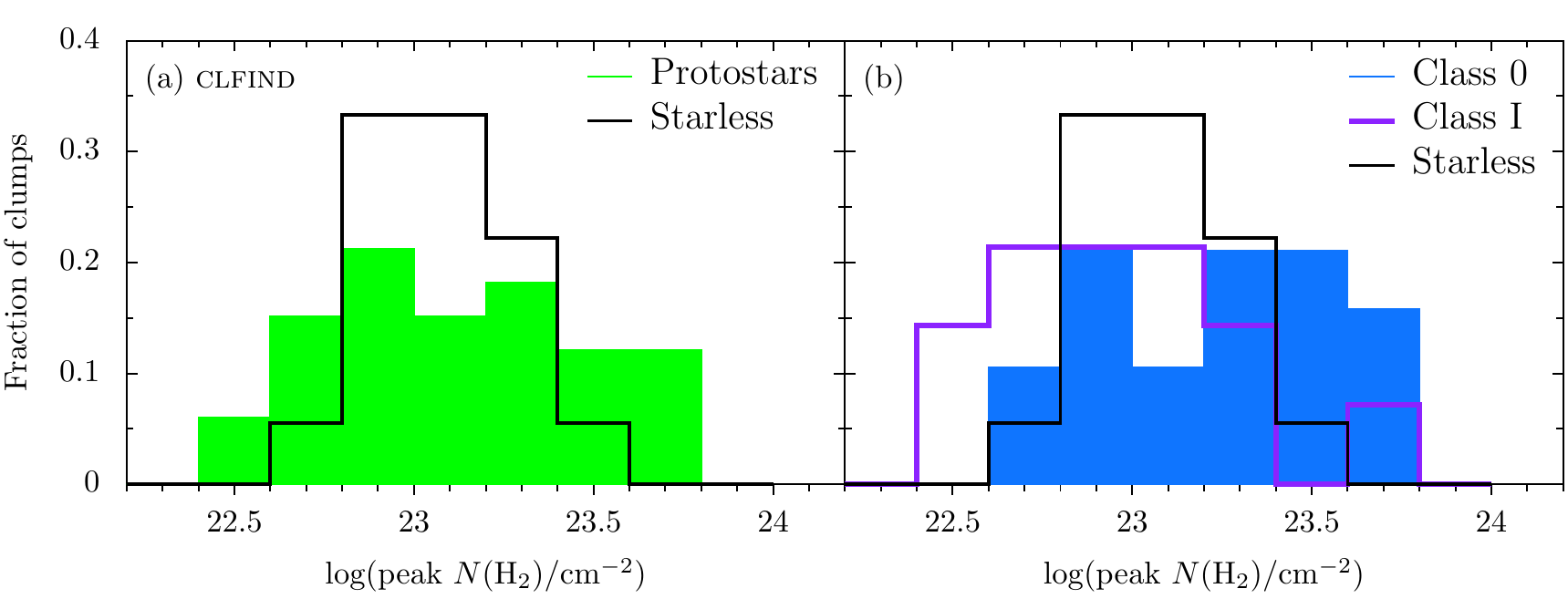}
\includegraphics[width=0.47\textwidth]{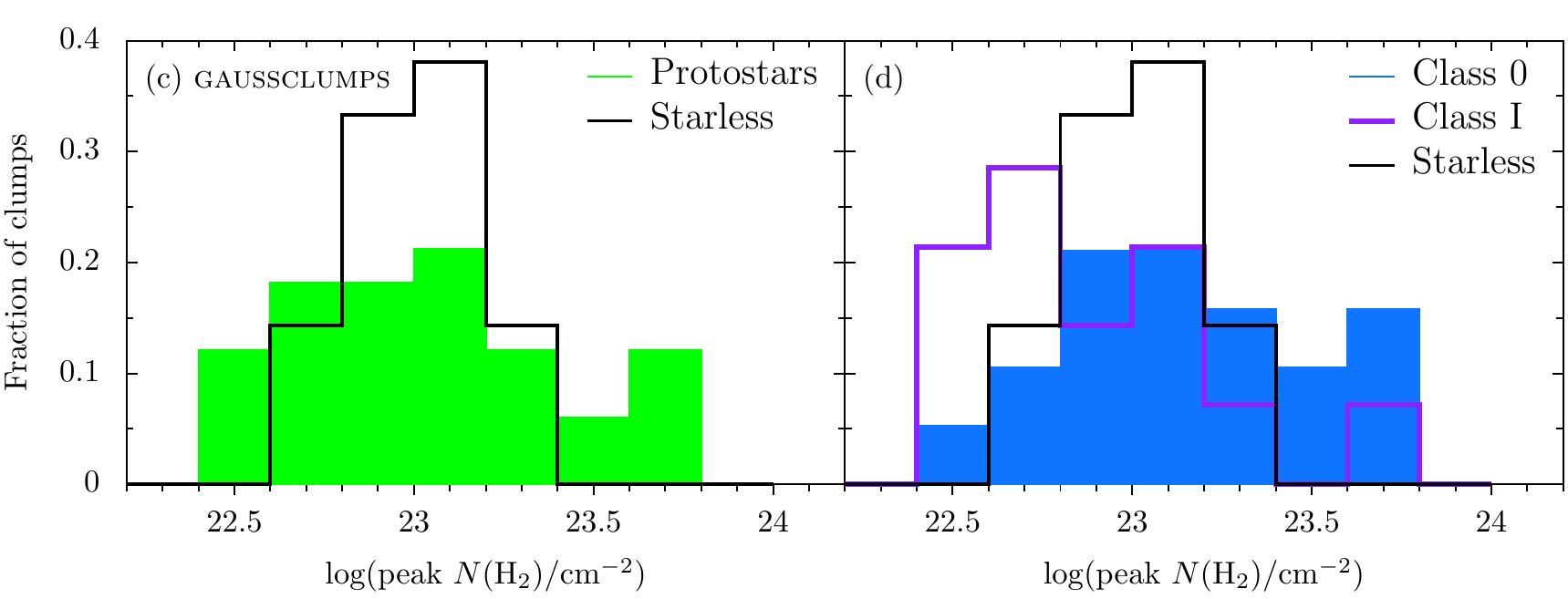}
\caption{Distribution of clump peak beam-averaged column densities for the clumps classified
  by age in the \citetalias{hatchell07a} catalogue for the \clfind\ (top, (a) and (b)) and \gclumps\ (bottom, (c) and (d)) populations. We assume a
  dust temperature of 10\,K and 15\,K for the starless and protostellar
  clumps respectively. Protostars have a similar average
  density to the starless cores but a wider distribution. The
  Class 0 sources are roughly the higher density half and Class Is
  the lower density half of the protostellar distribution. }
\label{fig:column_bytype}
\end{center}
\end{figure}

The peak beam-averaged column density of hydrogen, $N(\mathrm{H_2})$, depends on the peak 850\,$\mu$m
flux, $F_{850}$, via: \begin{equation} N(\mathrm{H_2}) =
  \frac{F_\nu}{\Omega_\mathrm{beam}\mu m_\mathrm{H_2} \kappa_\nu
    B_\nu(T_\mathrm{D})} \end{equation} where $\Omega_\mathrm{beam}$
is the solid angle subtended by
the beam, $m_\mathrm{H_2}$ the mass of a hydrogen molecule and
$\mu=1.4$ the mean molecular weight per H$_2$ molecule, assuming 5
H$_2$ molecules for every He. For the values
assumed in this paper: \begin{eqnarray} N(\mathrm{H_2}) = 7.62\times
  10^{22} \left(
  \frac{F_{850}}{\mathrm{Jy\,beam^{-1}}} \right) \left[
    \exp \left( 1.4 \left( \frac{12\,\mathrm{K}}{T_\mathrm{D}} \right)
    \right) -1 \right] \nonumber \\ \!\!\!\!\!\!\!\!\!\!\! \times  \left(
  \frac{\kappa_{850}}{0.012\,\mathrm{cm^2\,g^{-1}}}
  \right)^{-1}\,\mathrm{cm^{-2}.}  \label{eqn:N_dust} \end{eqnarray} We take $T_\mathrm{D}$ to be 15\,K for protostars, 10\,K
for starless agglomerations and  12\,K otherwise. Figs.\ \ref{fig:column_byregion} and
\ref{fig:column_bytype} are histograms of the peak clump column
density, whereas values for the individual populations are listed in
Tabs.\ \ref{tab:overallproperties_clfind} and
\ref{tab:overallproperties_gclumps}. These are necessarily population upper limits since lower density clumps are
below our sensitivity threshold. For the first time the \clfind\ and
\gclumps\ distributions qualitatively resemble one another,
with most clumps falling between 4 and $6.5\times 10^{22}$\,cm$^{-2}$, in the
second bin. Their mean log column densities are also close, slightly higher than the most
frequent bin at $(1.40\pm 0.18)\times 10^{23}$\,cm$^{-2}$ for
\clfind\ and $(1.11\pm 0.12)\times 10^{23}$\,cm$^{-2}$
for \gclumps. The distributions' regional variation does not seem
to be significant, with only L1455 (with the fewest clumps) showing markedly different behaviour. The breakdown by
age is more significant. Similar behaviour is seen for
the two algorithms, the protostars and starless cores have similar
means, $(2.6\pm 0.4)\times 10^{23}$ and $(1.00\pm 0.15)\times 10^{23}$\,cm$^{-2}$ respectively for
\clfind\ alongside $(2.3\pm 0.3)\times 10^{23}$ and
$(0.90\pm 0.09)\times 10^{23}$\,cm$^{-2}$ for \gclumps\, with the
protostellar sources having much wider
distributions. When the protostars are split into Class 0 and I
sources the lower densities are mainly
Class I sources (means of $(9\pm 2)\times 10^{22}$ and $(7.2\pm 1.5)\times
10^{22}$\,cm$^{-2}$ for \clfind\ and \gclumps) with
the higher Class 0 sources (means of $(3.4\pm 0.5)\times 10^{23}$ and
$(2.7\pm 0.3)\times
10^{23}$\,cm$^{-2}$ for \clfind\ and
\gclumps). From the Class 0/I definition, we have the
simple expectation that the more evolved Class I cores have accreted more
of their mass on to the central object. So there are lower densities
in Class I envelopes, particularly given their smaller or comparable
size to Class 0 objects.
 
\subsection{Mass versus size relationship} \label{sec:mversusr_scubaclumps}

\begin{figure}
\begin{center}
\includegraphics[width=0.47\textwidth]{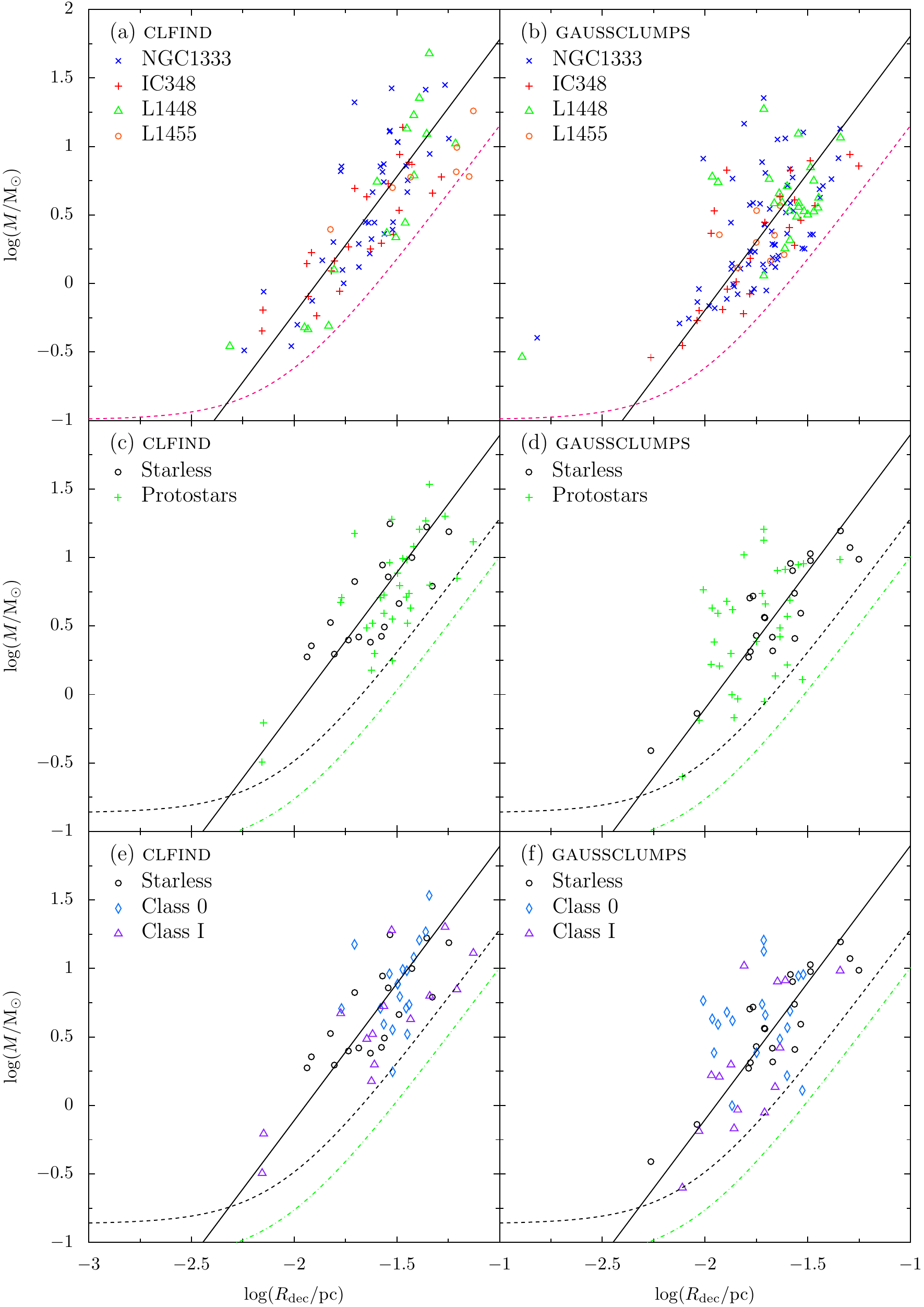}
\caption{Mass versus radius relation for the dust clumps in Perseus. The radius
  plotted, $R_\mathrm{dec}$, has been deconvolved with the beam (see Section
  \ref{sec:clump_sizes}). The plots (all on the same scales) are for
  the \clfind\
  (left panels, (a), (c) and (e)) and \gclumps\
  (right panels, (b), (d) and (f)) populations,
  differentiated by region (top panels, (a) and (b)) or by
  source type from the classifications of \citetalias{hatchell07a} (the
  rest). Lines are drawn at various completeness limits for constant \emph{observed} 
  surface brightness i.e.\ with $M\propto
  R^2$, where $R$ in this case refers to the \emph{observed} clump radius (not deconvolved with the beam
  size). The limits correspond to $4\sigma_\mathrm{rms}$ detections above the noise for 12\,K clumps
  (dashed pink), 10\,K starless cores (dashed black) and
  15\,K protostars (dot-dashed green). Solid black lines mark the
  best-fitting straight lines of the form $M\propto R^2$ (i.e.\ constant surface brightness), either fitted to the whole population ((a) and (b)) or just the
  starless clumps (plots (c) to (f)).}
\label{fig:massversusradius}
\end{center}
\end{figure}

Fig.\ \ref{fig:massversusradius} illustrates the mass-radius relation
for the clumps. A least-squares fit to the whole
population yields a straight line with a gradient of $1.7\pm 0.1$ for
\clfind\ and $1.2\pm 0.2$ for \gclumps. The slope is not
well constrained; the constant surface brightness scaling, $M\propto
R^2$, provides a good fit by eye. The scaling implies a surface brightness, $\Sigma(\mathrm{H_2}) = 1920$\,\msun\,pc$^{-2}$ and $\Sigma(\mathrm{H_2}) =
2470$\,\msun\,pc$^{-2}$ for the \clfind\ and \gclumps\ sources
respectively. These are around ten times higher than the average for
molecular clouds as a whole, $\bar{\Sigma}(\mathrm{H_2}) =
170$\,\msun\,pc$^{-2}$ \citep{solomon87} and four times the largest
$\Sigma$ found by \citeauthor{solomon87}, $\sim$500
\citep{heyer08}. This implies there is a large density contrast
between the clumps and their surrounding medium. 

\citet{enoch08} report distinct mass-radius relationships for protostellar and starless cores in
Perseus, with the protostars deviating from the tight $M\propto
R^2$ correlation shown by their starless counterparts. They calculate
source sizes by fitting an elliptical Gaussian (similar to
\gclumps\ but at previously identified peaks) and
masses by integrating the flux within a fixed circular aperture. Our
results are slightly different depending on which algorithm is
considered. The \clfind\ relations are very close
to $M\propto R^2$ for whichever class of object, whereas for
\gclumps, the scatter is greater around $M\propto R^2$,
with a somewhat greater range of masses for a narrower range of
sizes in the protostars. This latter observation loosely ties in with the simple suggestion of \citet{enoch08} that
the present protostellar population in Perseus evolved from prestellar cores which
decreased in size and density at constant mass until collapse and
protostellar formation. We are only detecting the
upper envelope of the mass-radius relationship, the bulk of the cores
may exist below the completeness limit. However, the small core
numbers in-between the completeness limits for 10 and 15\,K cores suggest
that we are not missing very many protostellar objects. 

In summary, many of the previously reported properties of
submillimetre dust clumps depend on the
method of identification and care has to be taken to properly account
for their biases. For instance, a simple evolutionary model with cores
becoming smaller and denser with time would fit the
\gclumps\ radii data with
$R(\mathrm{Starless})>R(\mathrm{Class~0})>R(\mathrm{Class~I})$.
However, \clfind\ data are contradictory:
$R(\mathrm{Starless})<R(\mathrm{Class~0})\approx
R(\mathrm{Class~I})$. Other patterns are more robust -- the protostellar and starless populations are
similarly elongated and the peak column densities of starless and
protostellar clumps are close on average with the Class 0 sources
occupying the upper half of the protostars' distribution and the Class
Is the lower. Larson's Law for constant surface brightness clumps,
$M\propto R^2$, is obeyed for both algorithms' clumps with more
scatter in the protostellar population of \gclumps.

\section{The clump mass distribution} \label{sec:cmd}

Every realistic model of star formation must match observations with the
fundamental test being whether it reproduces the stellar IMF. Interpretations
of CMDs are complicated by a number of
considerations, however appealing a direct one-to-one mapping
may seem (see e.g.\ \citealp*{andre08}):
\begin{enumerate}
\item \textbf{Form of the distribution.} The differential form of
  the mass distribution, though easier to
  interpret, is inaccurate when the number of sources is small
  ($\la 100$), with a cumulative distribution preferred
  \citep{reid06}.
\item \textbf{Dust properties.} Core masses from continuum measurements are based on uncertain
  assumptions about the dust itself (i.e.\ opacity and temperature).
\item \textbf{Completeness.} Sensitivity limits have the same
  effect as a real turn-over at lower masses, although in e.g.\ \citet{nutter07}, 
  a turn-over significantly higher than the completeness
  limit is claimed.  
\item \textbf{Nature of the objects.} Many submillimetre cores without
  a central object will \emph{not} form stars. Molecular-line
  observations can be useful in discriminating such \emph{unbound}
  objects from their \emph{bound} brethren (e.g.\ Curtis \& Richer, in
  prep.).  
\item \textbf{Multiplicity.} Some cores will 
  form multiple objects, indeed higher resolution observations often break
  cores down into smaller constituents. Furthermore, multiple systems can
  form after the prestellar phase during collapse (e.g.\ \citealp{goodwin07}). 
\item \textbf{Efficiency.} Typically a constant star-forming
  efficiency, i.e.\ how effectively a core turns its mass into
  stellar material, is invoked to map the CMD on to the IMF. If this efficiency is mass dependent then the mapping can be more
  complicated \citep{swift08}.
\item \textbf{Timescales.} \citet*{clark07} note that if core lifetimes depend on mass then the shape of
  the observed CMD will differ from the true, underlying distribution. 
\end{enumerate}

We have already mentioned the striking similarity between the CMD
and the stellar IMF and this section is a critical
examination of this relationship. To begin, we examine the
standard CMD for the SCUBA population, comparing it to previous work and looking at
regional and age variations. Finally we will
look at some of the distortions that prevent a simple mapping of the
CMD on to the IMF.

\subsection{Form of the mass function} \label{sec:cmd_form}

For the initial analysis of the entire region, we mainly use a
differential mass function, since its
interpretation is more straightforward. For clump numbers fewer than 100, where the arbitrariness
of binning becomes important, the cumulative form is preferable
\citep{reid06}. We fit the differential CMD using the following functional
forms, the most simple being a single power law: \begin{equation} \frac{\mathrm{d}N}{\mathrm{d}M} = AM^\alpha
\end{equation} with $\mathrm{d}N$, the number of objects in a mass
bin of width $\mathrm{d}M$, $\alpha$ the power-law exponent and $A$ a
constant. Frequently, when a turn-over is observed (at break mass, $M_\mathrm{break}$), a broken power law is
fitted with two components:

\begin{equation} \frac{\mathrm{d}N}{\mathrm{d}M}= \left\{ \begin{array}{l l}
A(M_\mathrm{break})^{\alpha_\mathrm{high}-\alpha_\mathrm{low}}M^{\alpha_\mathrm{low}} & \quad \mbox{for $M<M_\rmn{break}$,}\\
AM^{\alpha_\mathrm{high}} & \quad \mbox{for $M\ge M_\rmn{break}$}
  \end{array} \right. \end{equation} with $\alpha_\mathrm{high}$ and $\alpha_\mathrm{low}$
 the power law exponents above and below $M_\mathrm{break}$
 respectively. Additionally, we fit a log-normal distribution to the differential CMD: \begin{equation} \frac{\mathrm{d}N}{\mathrm{d}M} =
 \frac{A}{M\sigma} \exp \left[ - \frac{(\ln M - \mu)^2}{2\sigma ^2}
 \right] \label{eqn:lognormal}\end{equation} where $\sigma$ is the width, $\mu$ is related
 to the characteristic mass and $A$ a normalization constant. 

It is normally assumed that the
cumulative form of the CMD is also a power law
differing in its exponent from $\alpha$ by unity: \begin{equation}
  N(>M) = \int_M^\infty \frac{\mathrm{d}N}{\mathrm{d}M}\mathrm{d}M = -
  \frac{A}{\alpha+1}M^{\alpha+1} \label{eqn:1pl} \end{equation} for
$\alpha < -1$. In reality the
sum is not to infinity but some fixed mass, $M_\mathrm{max}$,
resulting from either an actual cut-off or sampling limit. Its shape is then modified: \begin{equation}
  N(>M) =
  -\frac{A}{\alpha+1}M^{\alpha+1}+\frac{A}{\alpha+1}{M_\mathrm{max}}^{\alpha+1} \label{eqn:1plmm} \end{equation} This modification can introduce significant curvature at shallow slopes, which could easily be confused with the need to fit a broken power law.

\subsection{Differential mass function} \label{sec:dmf}

\begin{figure}
\begin{center}
\includegraphics[width=0.47\textwidth]{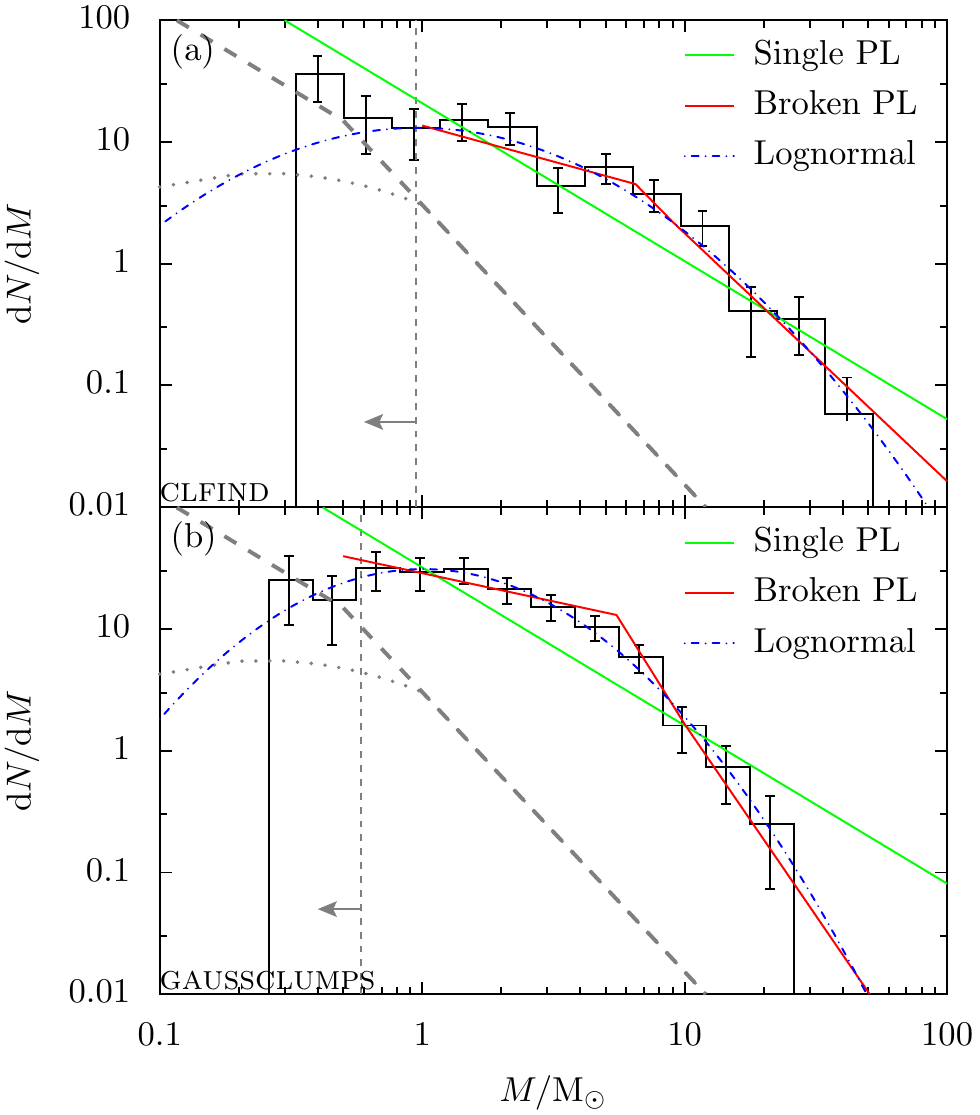}
\caption{Differential mass function of all the submillimetre clumps in
  Perseus across our four regions. The scales
  are the same for both plots as is the conversion to mass, with the standard
  assumptions including a constant clump temperature of 12\,K. The error-bars represent
  Poisson uncertainties. Vertical lines (dashed
  grey) mark the completeness limits (0.95 and 0.58\,M$_\odot$ 
  appropriate for \clfind\ and \gclumps\ respectively) calculated for
  averaged-sized sources of each algorithm population (22 and 17\,arcsec in turn). Two further lines mark the IMF for single stars
  (thick dashed grey, \citealp{kroupa01}) and
  multiple systems (dotted grey, \citealp{chabrier05}). Maximum
  likelihood fits to the mass function are plotted for various
  forms: a single power law (solid green), broken power law
  (solid red) and a log-normal distribution (dot-dashed blue).} 
\label{fig:dmf}
\end{center}
\end{figure}

The differential mass distribution is presented in Fig.\ \ref{fig:dmf}. The clumps of both algorithms span a similar range of
masses, with those from \gclumps\ extending to lower masses
and cut-off at a smaller maximum mass: 0.3  to
47.8\,\msun\ with an average, $\langle M \rangle = 6.4 $\,\msun\
for \clfind\ and  0.3 to 22.6 with $\langle M \rangle =
3.8$\,\msun\ for \gclumps. Maximum-likelihood fits of a power law to the
\clfind\ data (above the sample average completeness
limit as for all our fits), has an exponent, $\alpha = -1.5 \pm 0.1$, and reduced
chi-squared, $\tilde{\chi}^2=4.0$. A broken power law provides a better fit ($\tilde{\chi}^2=0.9$), with $\alpha_\mathrm{low} =
-0.6 \pm 0.1$ for $M<6.5$\,M$_\odot$ and $\alpha_\mathrm{high} =
-2.0 \pm 0.1$ for $M\ge 6.5$\,M$_\odot$. Additionally, a log-normal
fit has parameters $\mu=1.4\pm 0.2$ and $\sigma = 1.2 \pm 0.1$
with $\tilde{\chi}^2=1.9$. The \gclumps\ data are similarly poorly
fitted by a single power law ($\alpha = -1.30 \pm 0.08$,
$\tilde{\chi}^2=3.7$) but a broken power law is better
($\tilde{\chi}^2=0.5$) , with $\alpha_\mathrm{low} =
-0.46 \pm 0.07$ for $M<5.5$\,M$_\odot$ and $\alpha_\mathrm{high} =
-3.15 \pm 0.08$ for $M\ge 5.5$\,M$_\odot$. The best-fitting
distribution for the increased
curvature is log-normal with parameters $\mu=1.0\pm 0.1$ and $\sigma = 0.97 \pm 0.08$,
resulting in $\tilde{\chi}^2=0.3$. 

Our assumptions can affect the masses by up to a
factor of four: e.g.\ dust temperature, opacity and the
distance. Importantly, if these vary \emph{uniformly} for every clump, the \emph{shape} of the
distribution would be unchanged, merely translating it along the mass axis, altering the
break mass. Non-uniform variations will certainly
alter the shape. We will investigate temperature
variations between clumps shortly but we completely ignore opacity and
temperature changes across individual clumps which would
require radiative transfer modelling.

A CMD which
closely resembles the IMF implies that the final masses of stars
are set at the \emph{prestellar} stage during the core formation
process (e.g.\ \citealp{padoan02}). In other models of star formation e.g.\ competitive accretion,
we would not expect such a direct link with the earliest stages
(e.g.\ \citealp{bate05}). A more realistic scenario might be
intermediate between the two with dynamic interactions \emph{and} turbulence
playing a role \citep{klessen00}. The exact shape of the local IMF is still rather uncertain (e.g.\ \citealp{scalo05}), with
estimates of the power law slope from $\alpha = -2.3$ to --2.8 at
masses $\ga 1$\,M$_\odot$. For both of algorithms we find a break in the power law, significantly higher
than the average sample completeness mass, although it should
be stressed the sample is incomplete at every mass. At high masses we find a
steeper power law, slightly lower than the Salpeter value for
\clfind\ and higher for \gclumps. A steeper
slope implies a lack of high-mass objects (as is evident from the
spread of masses). \gclumps\ probably misses clump
mass at large distances from the clump centre, which can be found by
\clfind\ as it fits an arbitrary shape. The
log-normal CMD fits are somewhat wider than the
\citet{chabrier05} IMF (1.2 and 0.97 compared to 0.55) and their
characteristic masses are a factor of four or so larger (1.4 and 1.0
compared to 0.25\,M$_\odot$). 

Assuming it is constant, we can estimate the efficiency of star formation,
$\epsilon$, i.e.\ how much clump mass is converted into stellar,
via the ratios of characteristic masses in the CMD and IMF. We compare the CMD break to the same in the IMF, ($\sim$0.5\,\msun, \citealp{kroupa01}) yielding
$\epsilon=0.08$ and 0.09 for \clfind\ and \gclumps\
respectively. These are considerably smaller than the accepted value
of 0.3 for embedded clusters \citep{lada03}. Analysing all archived SCUBA
data, \citet{nutter07} and \citet*{simpson08} find smaller $M_\mathrm{break}$ of 2.4\,\msun\ in Orion and
2.0\,\msun\ in L1688 respectively. The different values suggest the efficiency depends on
environment. The break mass can be heavily affected by any
assumptions so we must exercise caution when making comparisons
between different clouds. For example, the Perseus survey of \citet{enoch06} finds the
break at 2.5\,\msun, but if we scale our results to match their
different assumptions about the dust opacity and clump temperatures,
we find our break mass is reduced to $3.25$\,\msun\ for \clfind\ and
similarly 2.25\,\msun\ for \gclumps. Correspondingly,
$\epsilon$ increases to 0.15 and 0.22 respectively and
there is less difference between clouds. Nevertheless the lower values of $\epsilon$ than
in clusters are probably because many cores will form multiple
stars breaking the one-to-one mapping with the IMF (see
\citealp{goodwin08,swift08}). 

A number of further caveats complicate any simple interpretation of
the CMD. If higher mass clumps form more stars
than lower mass ones, the CMD will be shallower than the IMF arising from it, as we found for \clfind. Additionally,
a clump's current mass may not be representative of the mass available for
accretion over its lifetime or the star formation efficiency may vary
with mass (see \citealp{hatchell08}).

\begin{figure}
\begin{center}
\includegraphics[width=0.47\textwidth]{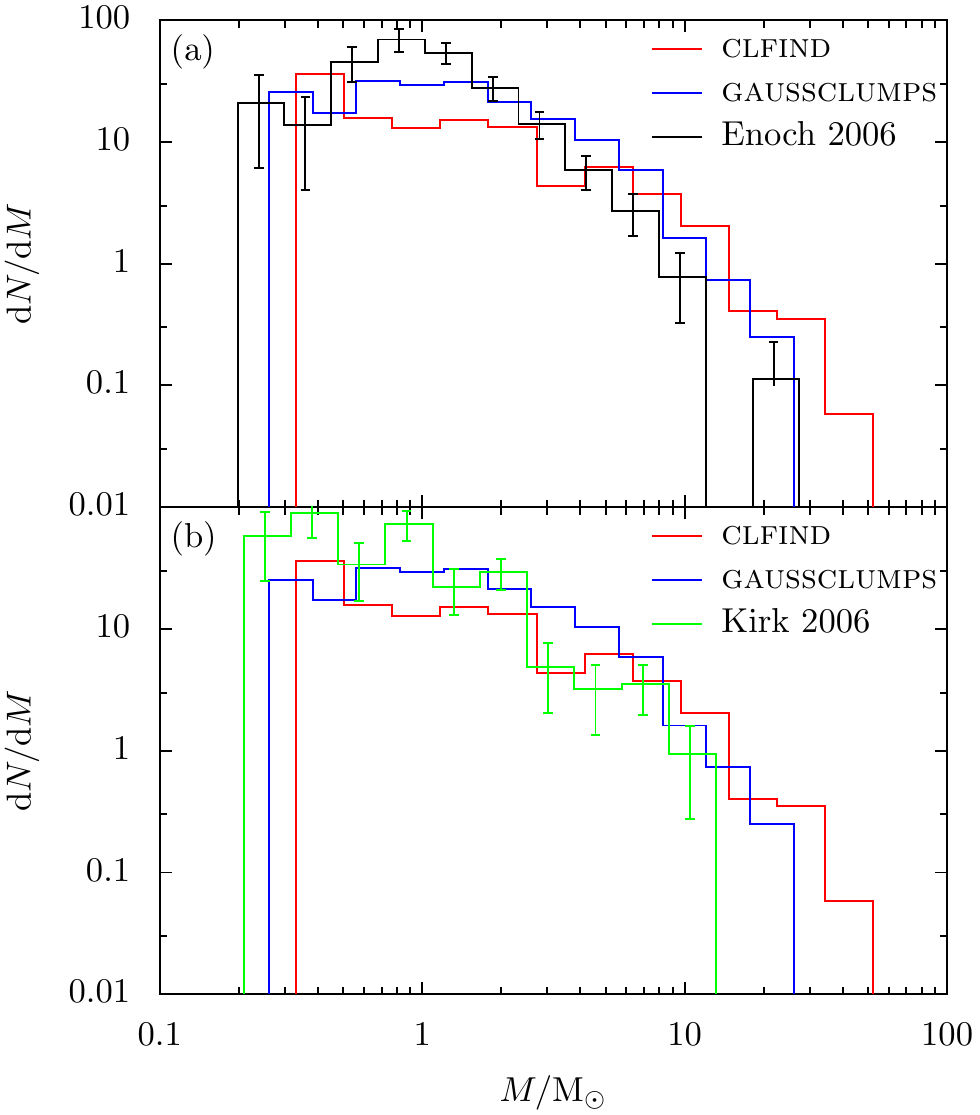}
\caption{Comparison of our differential mass functions, found with \clfind\ (red) and \gclumps\ (blue)
  to previous work. We plot distributions derived by \citet{enoch06} (top panel,
  black) and \citet{kirk06} (bottom panel, green), which correspond
  closely in shape to our mass functions. \citet{enoch06} surveyed
  7.5\,deg$^2$ of Perseus at 1.1\,mm
  with Bolocam on the CSO, whereas \citet{kirk06} used the data
  presented by \citet{hatchell05} combined with further independent
  SCUBA observations to yield a 3.5\,deg$^2$ survey. }
\label{fig:dmf_comparison}
\end{center}
\end{figure}

Other groups have measured the CMD in Perseus. The
widest survey used Bolocam \citep{enoch06} which we plot alongside
our results and the complementary SCUBA analysis of
\citet{kirk06} in Fig.\ \ref{fig:dmf_comparison}. \citet{enoch06} find 122 clumps, using a bespoke
algorithm with a more
stringent detection limit, a 5$\sigma_\mathrm{rms}$ peak above the
noise ($\sigma_\mathrm{rms}=15$\,mJy\,beam$^{-1}$) and larger beam (31\,arcsec FWHM). Most sources in Perseus lie in five
clusters, four of which we analysed, so our comparable numbers are
not surprising. \citet{kirk06} find
considerably fewer sources, only 58, even though they incorporated
these data. They used their own data reduction technique, convolved to a degraded map
resolution of 19.9\,arcsec. The resulting maps had higher
noise and they used the more stringent 5$\sigma_\mathrm{rms}$ peak flux
criterion with \clfind, which probably explains the difference. 

\citet{enoch06} find a similar spread of masses (0.2 to 25.6\,M$_\odot$), reaching out to lower 
masses due to their better sensitivity. \citet{kirk06}
have a narrower range (0.2 to 9.8\,M$_\odot$), the absence of high mass clumps might arise as they have lost sensitivity to low
    density material with their higher noise and degraded
    resolution. Both groups adopted different assumptions; to match \citet{enoch06} and
    \citet{kirk06}, our data would need to be multiplied by 0.5 and 0.9
    respectively. \citet{enoch06} find their data is best fitted by a
    broken power law: $\alpha_\mathrm{low} =
-1.3 \pm 0.3$ for $M<2.5$\,M$_\odot$ and $\alpha_\mathrm{high} =
-2.6 \pm 0.3$ for $M\ge 2.5$\,M$_\odot$. This is slightly steeper than for
both of our algorithms at low masses and almost precisely
intermediate between \clfind\ and \gclumps\
at high. The \citet{kirk06} data are best fitted by a single power
law that is quite shallow, $\alpha=-1.5\pm 0.2$ but since there is a
small number of clumps a cumulative mass distribution is more appropriate.
 
\subsection{Regional variations} \label{sec:byregion}

\begin{figure}
\begin{center}
\includegraphics[width=0.47\textwidth]{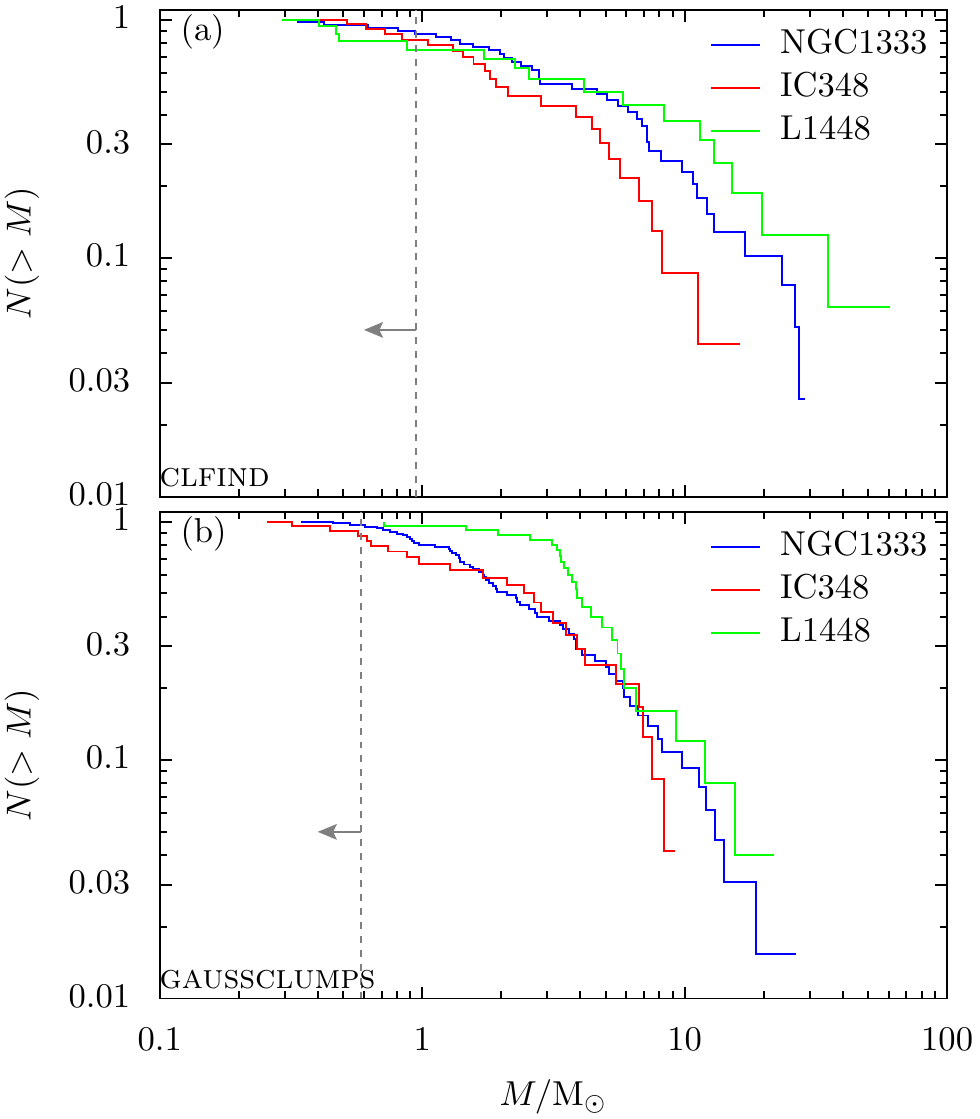}
\caption{Normalized cumulative mass distributions across three
  of our regions for the \clfind\ (top, (a)) and \gclumps\ (bottom,
  (b)) populations. The 4$\sigma_\mathrm{rms}$ completeness masses for a
  source of the average sample size are also plotted (dotted grey). }
\label{fig:cmd_byregion}
\end{center}
\end{figure}

In Fig.\ \ref{fig:cmd_byregion}, we plot cumulative CMDs for
three of our regions in Perseus, L1455 has been excluded because it has
a small number of clumps. We use the cumulative form as there are only a small
number of cores in each region. The
outcomes of maximum-likelihood single and double power law fitting to
the CMD with K-S tests to differentiate between the fits are
presented in Tabs.\ \ref{table:byregion_fits} and
\ref{table:byregion_prob}. Clumps in L1448 are more
massive than those in NGC~1333, which are more massive in turn than
those in IC348. Returning to the various properties of the regions in Tab.\ \ref{table:region_diagnostics}, this implies that less clustered regions have more
massive clumps (as we may expect from an analogous argument to that
employed for the size) and regions that have higher protostellar
fractions are more massive. The best-fitting models are typically single
power laws with the maximum mass taken properly into account for
\clfind\ or broken power laws for \gclumps. The high-mass slope appears to be greater for IC348
than NGC~1333 or L1448, although the slope is very sensitive to the
exact break mass and fitted model. This is not clear from the fits for
\gclumps\ until the last points are looked at on the
plot. Therefore there is a lack of high-mass
objects in the IC348, which is likely to be caused by its
young population, containing the largest proportion of starless cores. For instance, if starless
cores are systematically colder than the assumed temperature
(12\,K), we would underestimate their masses. Additionally, their masses may be inherently different to protostars as
we will explore shortly. The distribution in L1448 varies markedly
between the algorithms with the \gclumps\ slope being significantly
steeper than the one for \clfind. A shallow slope could be explained by
its high protostellar fraction, if they are over-estimated in mass by
the 12\,K assumption or are intrinsically
heavier. However, the
steep \gclumps\ slope for L1448 would bring us to the opposite
conclusion.

\begin{table}
\caption{Best-fitting parameters for various models of the cumulative CMD split up
  by region. The models that are best-fitting to the data for each
  region are marked with an asterisk (evaluated using the K-S $D$
  values presented in Tab.\ \ref{table:byregion_prob}).}
\begin{scriptsize}
\begin{tabular}{lcccc}
\hline
Region & 1PL~$^\mathrm{a}$ & 1PLMM~$^\mathrm{b}$& \multicolumn{2}{c}{2PL~$^\mathrm{c}$} \\
$\backslash$Model & $\alpha$ & $\alpha$ & $\alpha_\mathrm{low}$ &
$\alpha_\mathrm{high}$ \\
\hline
\multicolumn{5}{l}{\clfind:} \\
NGC~1333&$-1.58\pm 0.03$& $-1.089\pm 0.003$* &$-1.47\pm 0.02$&$-2.39\pm 0.04$\\
IC348&$-1.79\pm 0.05$&$-1.15\pm 0.02$* &$-1.69\pm 0.04$&$-2.81\pm 0.05$
\\
L1448&$-1.46\pm 0.06$* & no fit &$-1.32\pm 0.01$&$-2.39\pm 0.06$ \\
\hline
\multicolumn{5}{l}{\gclumps:}\\
NGC~1333&$-1.68\pm 0.02$& $-1.43\pm 0.04$ & $-1.60\pm 0.02$&$-2.55\pm
0.04$* \\
IC348&$-1.56\pm 0.08$& $-1.0069\pm 0.0003$ & $-1.41\pm 0.04$ &
$-2.43\pm 0.10$* \\
L1448&$-1.72\pm 0.07$& $-1.23\pm 0.02$ & $-1.13\pm 0.01$* & $-2.89\pm
0.04$ \\
\hline
\end{tabular}
\end{scriptsize}

$^\mathrm{a}$~Single power law fit (1PL), as in Eq.\ \ref{eqn:1pl} ($\alpha$ everywhere corresponds to the gradient of an
  equivalent differential CMD).\\ 
$^\mathrm{b}$~Single power law fit taking into account
  the finite upper limit of integration (i.e.\ maximum mass, 1PLMM), as in Eq.\ \ref{eqn:1plmm}. \\
$^\mathrm{c}$~Broken power law fit (2PL).
\label{table:byregion_fits}
\end{table}

\begin{table}
\caption{K-S $D$
  values for the various fits of the cumulative CMD by region. For the
  different model names see Tab.\ \ref{table:byregion_fits}. An asterisk marks
  the best-fitting model for each region.}
\begin{tabular}{llll}
\hline
Region \& algorithm $\backslash$Model & 1PL & 1PLMM & 2PL \\
\hline
NGC~1333, \clfind\ & 0.13 & 0.06* & 0.17 \\
IC348, \clfind\ & 0.09 & 0.06* & 0.12 \\
L1448, \clfind\ & 0.09* & -- & 0.18 \\
NGC~1333, \gclumps\ & 0.13 & 0.09 & 0.08* \\
IC348, \gclumps\ & 0.33 & 0.12 & 0.07* \\
L1448, \gclumps\ & 0.26 & 0.20* & 0.34 \\
\hline
\end{tabular}
\label{table:byregion_prob}
\end{table}

\subsection{Variations by source age} \label{sec:cmd_bytype}

An examination of how the CMD varies with source age
can lead to crucial insights into the star formation process. In a
trivial mapping on to the IMF, with a constant
star-forming efficiency, the shape at the \emph{prestellar} phase
may distinguish between models where
the mass is set by: (i) the initial fragmentation on to cores, (ii)
competitive accretion or (iii) protostellar mechanisms i.e.\ outflows. 

\begin{figure}
\begin{center}
\includegraphics[width=0.47\textwidth]{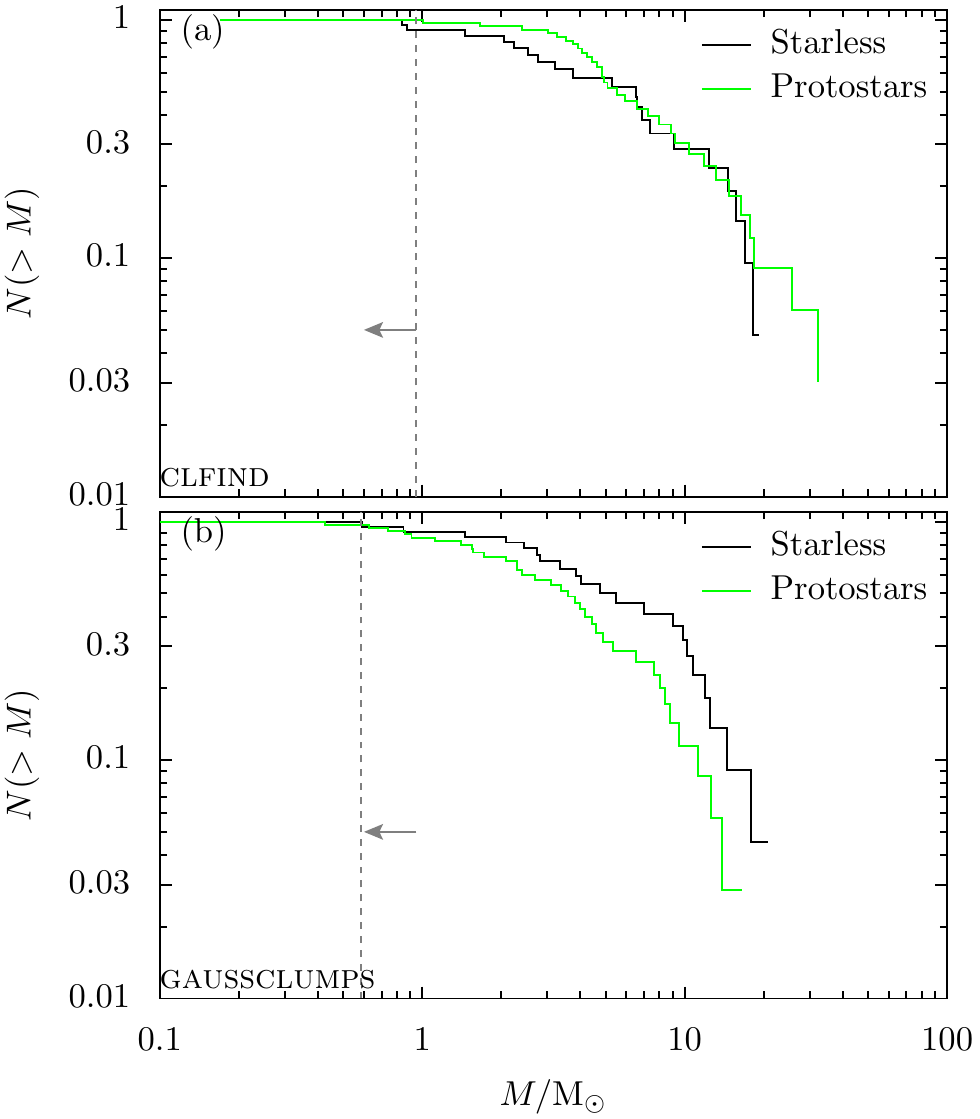}
\caption{Normalized
  cumulative mass distributions for protostars (green) and starless cores (black)
  as designated by \citetalias{hatchell07a} for the \clfind\ (top,
  (a)) and \gclumps\ (bottom, (b)) populations. The 4$\sigma_\mathrm{rms}$ completeness masses for a
  source of the average sample size are also shown (dotted
  grey). The masses were calculated
  assuming a dust temperature of 10\,K for the starless and 15\,K for
  the protostellar sources.}
\label{fig:cmd_byage}
\end{center}
\end{figure}

\begin{figure}
\begin{center}
\includegraphics[width=0.47\textwidth]{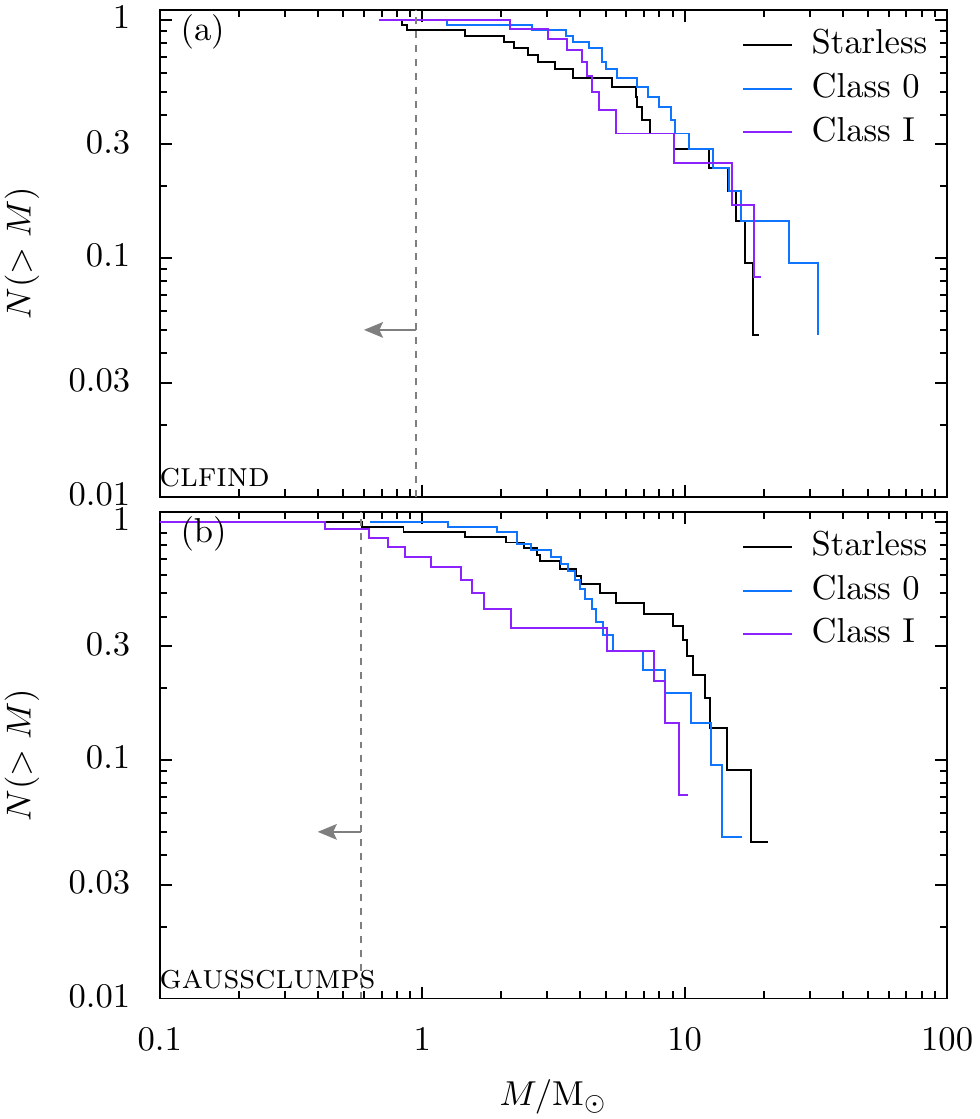}
\caption{Normalized
  cumulative mass distributions for starless
  cores (black), Class 0 (blue) and Class I
  (purple) protostars as designated by \citetalias{hatchell07a} for the \clfind\ (top,
  (a)) and \gclumps\ (bottom, (b)) populations. The 4$\sigma_\mathrm{rms}$ completeness masses for a
  source of the average sample size are also shown (dotted
  grey). The masses were calculated
  assuming a dust temperature of 10\,K for the starless and 15\,K for
  the Class 0/I sources.}
\label{fig:cmd_bysource}
\end{center}
\end{figure}

In Fig.\ \ref{fig:cmd_byage}, we plot the cumulative CMD for protostars
and starless cores with Fig.\ \ref{fig:cmd_bysource}
further breaking the protostars into Class 0 and Is. The
masses have been calculated in the usual way assuming isothermal
dust temperatures of 10\,K for
starless and 15\,K for protostellar cores. As in Section
\ref{sec:byregion}, we fit models to the starless and protostellar
cumulative CMDs (see Tabs.\ \ref{table:byage_fits} and \ref{table:byage_prob}). For \clfind, the protostellar CMD is wider, extending to higher
and lower masses than the starless equivalent. The situation for \gclumps\
is more complicated, with the protostellar distribution offset to
lower masses than the starless one. When the protostars are
further divided, the Class 0 sources have a flatter distribution,
containing all the high mass objects, than the Class I or starless
distributions with \clfind. The starless clumps are at lowest masses with
the Class Is in the middle. The same holds for the
\gclumps\ decomposition, except the Class I population
extends to lowest masses and the starless population out to the highest. 

\begin{table}
\caption{Best-fitting parameters for various models of the cumulative CMD split up
  by the source identifications from \citetalias{hatchell07a}. For the
  different model names see Tab.\ \ref{table:byregion_fits}. The models that are best-fitting to the data for each
  source population are marked with an asterisk (evaluated using the K-S $D$
  values presented in Tab.\ \ref{table:byage_prob}). }
\begin{scriptsize}
\begin{tabular}{lcccc}
\hline
Region & 1PL & 1PLMM & \multicolumn{2}{c}{2PL} \\
$\backslash$Model & $\alpha$ & $\alpha$ & $\alpha_\mathrm{low}$ &
$\alpha_\mathrm{high}$ \\
\hline
\multicolumn{5}{l}{\clfind:} \\
Starless&$-1.69\pm 0.04$& $-0.62\pm 0.04$*&$-1.57\pm 0.04$&$-5.17\pm 0.07$\\
Protostars&$-1.65\pm 0.04$&$-1.19\pm 0.01$*&$-1.53\pm 0.05$&$-2.71\pm 0.05$
\\
\hline
\multicolumn{5}{l}{\gclumps:}\\
Starless&$-1.52\pm 0.04$& $-0.68\pm 0.05$*&$-1.32\pm 0.05$&$-2.38\pm 0.04$\\
Protostars&$-1.59\pm 0.04$&$-0.69\pm 0.04$*&$-1.48\pm 0.05$&$-3.71\pm 0.04$
\\
\hline
\end{tabular}
\end{scriptsize}
\label{table:byage_fits}
\end{table}

\begin{table}
\caption{K-S $D$
  values for the various fits of the cumulative CMD split up
  by source identifications from \citetalias{hatchell07a}. For the
  different model names see Tab.\ \ref{table:byregion_fits}. An asterisk marks
  the best-fitting model for each category.}
\begin{tabular}{lccc}
\hline
Region \& algorithm $\backslash$Model & 1PL & 1PLMM & 2PL \\
\hline
Starless, \clfind\ & 0.14 & 0.10* & 0.19 \\
Protostars, \clfind\ & 0.35 & 0.16* & 0.18 \\
Starless, \gclumps\ & 0.23 & 0.06* & 0.09 \\
Protostars, \gclumps\ & 0.15 & 0.06* & 0.11 \\
\hline
\end{tabular}
\label{table:byage_prob}
\end{table}

This is the part of the picture described in \citetalias{hatchell07a} and
\citet{hatchell08}. For the same data analysed in a similar fashion with 
\clfind, they find comparable results: Class 0 sources are
more massive than Class Is and there is an excess of Class 0 protostars at
high masses relative to starless cores -- there is a
lack of massive starless cores. It is reassuring that the Class 0
sources are more massive than Class Is. The original definition
proposed sources to be: Class 0 if the ratio of envelope to
central object mass,  $M_\mathrm{env}/M_\star>1$ and Class I if
$M_\mathrm{env}/M_\star<1$ \citep*{andre93}. 

Our results are somewhat inconclusive. Single power law fits to the
distributions derived by the two algorithms yield similar slopes for protostellar and
starless sources. A power law fit, including the maximum mass term
(Eq.\ \ref{eqn:1plmm}), has a similar slope for both source types with \gclumps\ and a shallower slope for the
starless cores with \clfind. This results because the maximum
protostellar mass is considerably greater than the starless one, so the necessary curvature is
already present and the slope can be
shallower. A broken power law, however, has a steeper
slope for starless over protostellar cores at the high mass end for
\clfind\ and the reverse for \gclumps. 

All-in-all, the \gclumps\ distributions are very similar to
one another. The Class I population is less massive than the Class 0
and starless populations on average, which is again not
surprising given their definition. We expect more evolved sources to have accreted more
mass and thereby have less massive dust envelopes. The \clfind\
data support the view that the most massive objects are Class 0
protostars and there is a paucity of massive starless cores. This
could be explained by dust temperature variations or a mass-dependent
evolution of the clumps, with massive prestellar clumps evolving rapidly
into protostars, hence not as many are detected (see
\citealp{hatchell08}). However, if starless cores are not
gravitationally bound, then they might not go on to form stars, so
placing them in such an evolutionary sequence would be misleading. We
examine their dynamic stability in a forthcoming paper (Curtis \&
Richer, in prep.).

\subsection{Effect of varying clump dust temperatures}

\begin{figure}
\begin{center}
\includegraphics[width=0.47\textwidth]{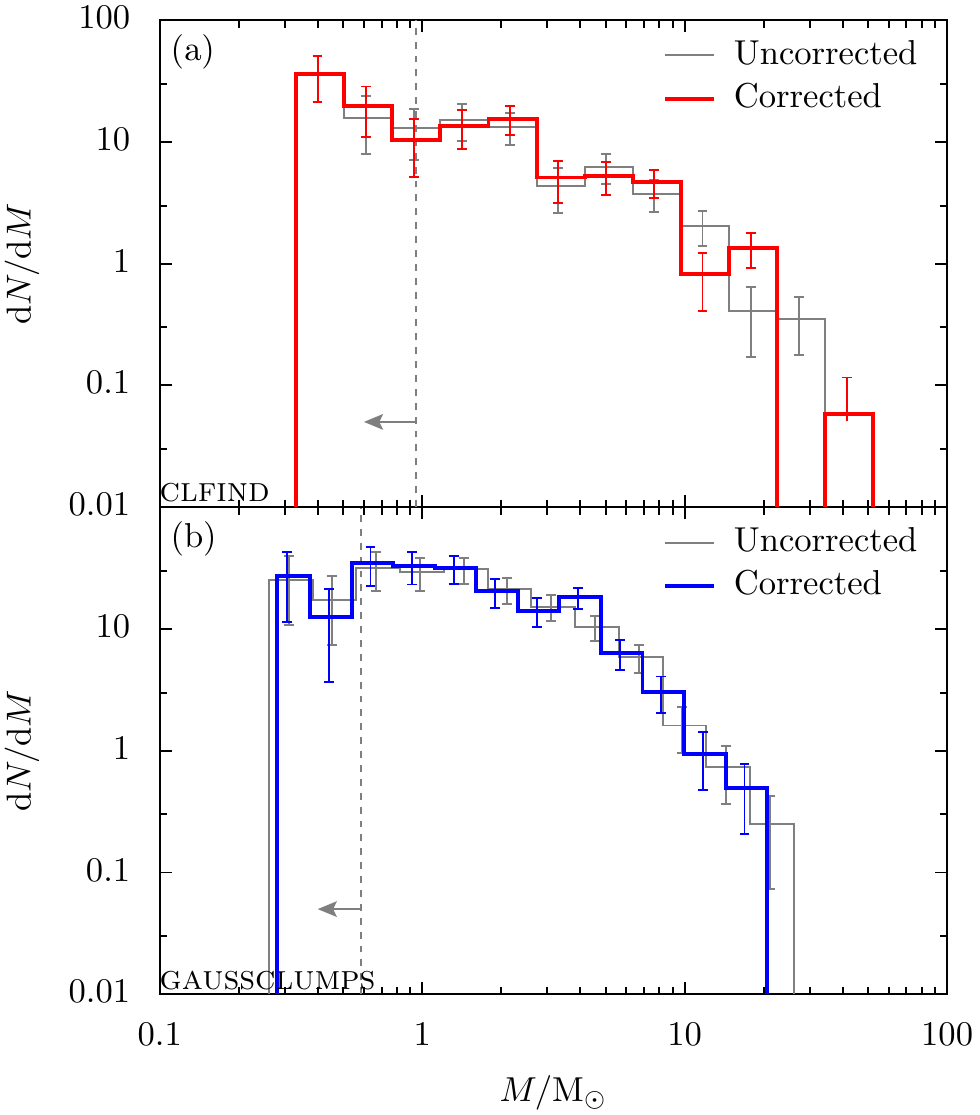}
\caption{Effect of varying clump temperatures
  on the mass distribution. The original constant-temperature distributions (grey)
  have been overlaid with ones calculated using NH$_3$ kinetic
  temperatures from the atlas of \citet{rosolowsky08} for the \clfind\
  (top, (a), red) and \gclumps\ (bottom, (b), blue) populations. The bars are standard Poisson errors on the
  bins and the 4$\sigma_\mathrm{rms}$ completeness masses for a source of average
  size are also shown (dashed grey).}
\label{fig:temp_correction}
\end{center}
\end{figure}

If the dust temperature varies between clumps then the
shape of the mass distribution will change. Recently, \citet{rosolowsky08} published NH$_3$ kinetic
temperatures, $T_\mathrm{k}$, towards 193 positions in Perseus using a
beam size of 31\,arcsec. If the gas and dust are well-coupled, typically occurring at high
densities, we would expect $T_\mathrm{D}\sim T_\mathrm{k}$. Of
these atlas positions, 50 coincide with our \clfind\
identifications (59~per cent of the clumps, $\bar{T}_\mathrm{k}=(12.9\pm 0.3)$\,K) and 49 with \gclumps\
sources (40~per cent,  $\bar{T}_\mathrm{k}=(12.6\pm 0.3)$\,K). In Fig.\ \ref{fig:temp_correction}, we plot the mass
distributions for the two algorithms using a varying dust temperature
equal to
$T_\mathrm{k}$ where it was coincident and 12\,K where not. The
distribution only changes marginally, indeed
\citet{enoch08} found the same for a similar analysis on the Bolocam data -- the
deviation from a power law actually \emph{decreased}. This is not
surprising since the average
temperature of the clumps is very similar to our assumed dust
temperature and their spread is small. 

\subsection{Effect of clump lifetimes}

\begin{figure}
\begin{center}
\includegraphics[width=0.47\textwidth]{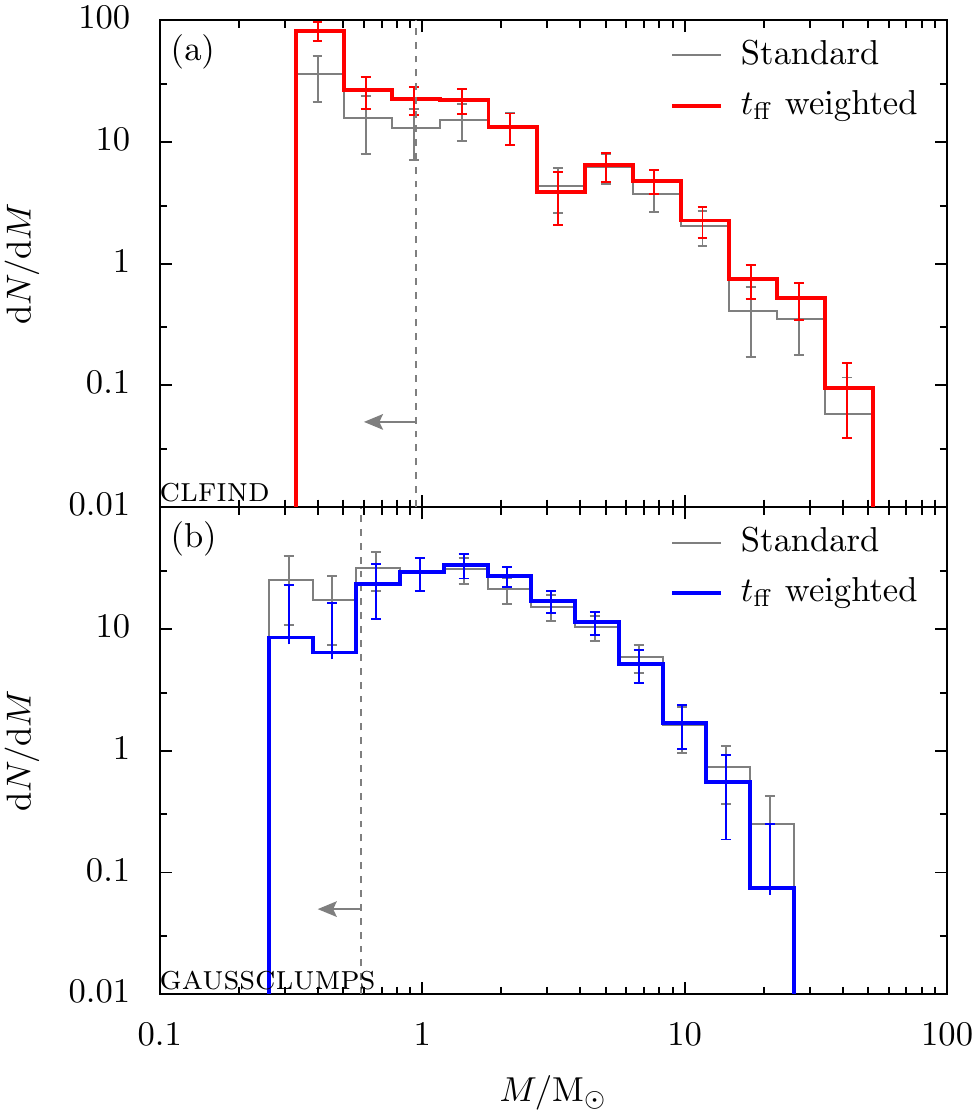}
\caption{Differential mass
  distribution weighted by the inverse of the clump free-fall timescale, $\langle t_\mathrm{ff}
\rangle/t_\mathrm{ff}$, with the original distributions
  in grey for the \clfind\
  (top, (a), red) and \gclumps\ (bottom, (b), blue) populations. The bars are standard Poisson errors on the
  bins and the 4$\sigma_\mathrm{rms}$ completeness masses for a source of average
  size are also shown (dashed grey).}
\label{fig:weighted_cmd}
\end{center}
\end{figure}

The observed mass distribution will not
be representative of the true distribution if the lifetime
of the cores is mass dependent, as we are more likely to detect
long-lived cores than short-lived ones. \citet{andre07} suggest a
\emph{weighted} mass function, with each core weighted according to
the inverse of it free-fall timescale ($\langle t_\mathrm{ff}
\rangle/t_\mathrm{ff}$, with $\langle t_\mathrm{ff}
\rangle$ the average core free-fall time). Applying such weighting to
the cores in Ophiuchus from \citet{motte98}, they find
the break in the
power law disappears and a single Salpeter-like power law is the best-fitting. In this section we attempt a similar weighting for our clumps. The
free-fall timescale of a homogeneous sphere
is: \begin{equation} t_\mathrm{ff} = \left( \frac{3\pi}{32G\rho}
  \right) ^{1/2} =\left( \frac{ \pi^2 R_\mathrm{dec}^3}{8GM_{850}}
  \right) ^{1/2} \end{equation} with $\rho$, the density
in the clump, which we estimate from previously derived quantities
as shown. We find free-fall timescales approximately in the range 10$^4$
to 10$^5$\,yrs. The resultant weighted distribution is shown in Fig.\
\ref{fig:weighted_cmd}. There is no apparent reduction in the
flattening of the distribution at low masses. The reasons for this may
be twofold. First, our estimates of the free-fall timescale may be
in error. Other groups estimate densities using the mass and
radius inside a fixed aperture, whereas our radii are
influenced by the whole of the clump. This will tend to
reduce the average density of large radius clumps, i.e.\ large mass
clumps, thereby increasing their free-fall timescales and reducing
their corresponding weight in the new distribution. This steepens the mass distribution at large
masses and should therefore not affect the presence of a break. Second,
there is, as other groups have noted, no relationship between mass
and $t_\mathrm{ff}$. If the break
in the CMD is related to differential clump lifetimes
then $t_\mathrm{ff}$ should in some way be related to the mass. 

\section{Summary} \label{sec:summary}

This paper detailed a new analysis of SCUBA 850\,\micron\ data towards
NGC~1333, IC348/HH211, L1448 and L1455 in the Perseus molecular cloud. We used two clump-finding algorithms, \clfind\ and \gclumps, to decompose the dust
column density into clumps of emission. \clfind\ located 85 clumps in
total, whilst \gclumps\ found 122. Once identified we divided the
clumps into protostellar (Class 0 and I) and starless sources by
pairing them with classifications in the catalogues of
\citetalias{hatchell07a}. 

This approach highlights trends in the clump properties that are highly or weakly dependent on
the clump-finding algorithm. Conclusions which hold regardless of the
technique include:
\begin{itemize}
\item \textbf{Shapes}. Cores are slightly elongated with a
  tendency to become more elliptical over time as expected from collapse models. 
\item \textbf{Peak column densities}. Protostars and starless
  cores have similar average peak column densities. The upper half of the protostellar
  distribution arises from Class 0 protostars and the lower from Class
  Is. This is expected from the class definitions and supports the
  classifications of \citetalias{hatchell07a}. Unsurprisingly, the masses of
  Class 0 clumps are also larger than Class Is. 
\item \textbf{Mass versus size}. Clumps are consistent with the $M
  \propto R^2$ relation, having average surface brightnesses some 4-10 times larger than typical giant molecular clouds.
\item \textbf{CMD}. The clump mass distribution resembles the IMF. The break mass is significantly higher than the average
  completeness mass and implies a star-forming efficiency of $\sim
  10$--20~per cent, probably reduced relative to embedded clusters
  due to multiplicity. 
\item \textbf{Regional variations}. The mass of clumps in each
  region follows:
  $M_\mathrm{clump}(\mathrm{L1448}) >
  M_\mathrm{clump}(\mathrm{NGC~1333}) >
  M_\mathrm{clump}(\mathrm{IC348})$. The high-mass slope of the {\sc
  CMD} is steeper for IC348 than L1448 with NGC~1333 intermediate. 
\item \textbf{CMD temperature and lifetime dependence}. Using
  individual temperatures for each clump or weighting according to
  their free-fall timescale does not change the shape of the CMD.
\end{itemize}

Other inferences are dependent on the clump identification procedure:
\begin{itemize}
\item \textbf{Sizes}. Protostars are either larger or smaller
  than prestellar cores with \clfind\ or \gclumps\ respectively. 
\item \textbf{Functional form of the CMD}. A log-normal CMD is
  best-fitting to the \gclumps\ data whereas a broken power law is
  best for \clfind. In any case a broken-power-law fit has different slopes for
  each algorithm, with the \gclumps\ high-mass slope steeper at $\alpha= -3.15\pm 0.08$ than $\alpha
  = -2.0\pm 0.1$ for \clfind. 
\item \textbf{A lack of massive starless cores?} A shortage of
  high-mass starless clumps relative to protostars was definitely present in the
  \clfind\ data (see also \citetalias{hatchell07a}; \citealp{hatchell08}). The same was not clear in
  the \gclumps\ data, where the overall spread in masses is
  smaller and the starless cores have larger sizes.   

\end{itemize}

\gclumps\ can only fit a strict shape so will not allocate flux to a clump
at large distances from the peak unlike \clfind. Therefore the
corresponding \gclumps\ sizes are smaller. The clumps sizes in clustered regions such as NGC~1333 are
smaller, as expected, for \clfind\ but not any
different for \gclumps, which is better at dealing with blended sources. The clump mass data are even more affected
by selection effects. The slope of the
starless and protostellar CMDs are very different for the two
algorithms. Furthermore, the lack of high-mass starless cores, previously observed
by \citetalias{hatchell07a} and implied by the shallow prestellar mass
function found by \citet{enoch08} is not apparent in the
\gclumps\ data. This would imply that either the shortage
is not real or \gclumps\ does not allocate enough flux to
high-mass clumps. Indeed, protostars which have more irregular, disrupted
envelopes due to e.g.\ outflows are perhaps worse affected by the
regular profiles used by \gclumps. This would cause a lack of high-mass
protostellar clumps, which might explain the \gclumps\ results.

Fundamentally, we can never know the real population of clumps
underlying any dataset. However, certain evolutionary
trends \emph{must} be present in any clump decomposition for it to be a believable
representation of this true population. One clear example of this is
the evolution of core size over time. Every simple model of collapse
naturally has cores decreasing in size with time. This trend is \emph{not}
apparent for the \clfind\ clumps. In addition, particular \clfind\ clumps
incorporate material that seems equally likely to go on to form part
of their final mass or that of their neighbours. Therefore, treating
the properties of a \clfind\ clump as representative of
gravitationally-bound material in a single core is
probably highly misleading. Of course, we may need to tweak the
\clfind\ parameters, increasing the lowest contour level and
decreasing the contour separation, but any method which necessarily prevents
overlap in a highly-cluster region, such as Perseus, will struggle to
find the true source population. Since one is normally interested in the material that will go on to form the final
star, \gclumps\ would seem the more representative algorithm choice.

Analysing the clumps found by two of the most popular clump-finding
routines side-by-side, underlines the caution that one must exercise
when comparing results from distinct studies on similar datasets in
the \textit{same} region. Different methods can produce very different results
from the same data, providing clump populations that are not directly
comparable. Future continuum and molecular-line surveys will need to
carefully design their source extraction procedures to ensure objects
are found at the desired length-scale without simply using an
automated algorithm blindly.   

\section{Acknowledgments}

EIC thanks the Science and Technology Facilities Council (STFC) for studentship support while carrying out this
work. The authors thank Jane Buckle and Gary Fuller for carefully
reading and suggesting improvements to earlier versions of this
paper and the referee, Derek Ward-Thompson, for his helpful
suggestions that significantly improved the final version. The JCMT is operated by The Joint Astronomy
Centre (JAC) on behalf of the STFC of
the United Kingdom, the Netherlands Organisation for Scientific
Research and the National Research Council (NRC) of Canada. We acknowledge the data analysis
facilities provided by the Starlink Project which is maintained by JAC
with support from STFC. This research used the facilities of the Canadian
Astronomy Data Centre operated by the NRC with the support of the Canadian Space Agency.

\appendix
\section{Clump catalogues} \label{appendix:catalogues}

\begin{table*}
\begin{flushleft}
\caption{Properties of SCUBA clumps found with \clfind. The full version of this table is available as Supplementary Material to the online version of this article.} 
\label{tab:clfind_detections}
\begin{tabular}{c c c c c c c c}
\hline
Sub-region  & Clump  & RA~$^\mathrm{a}$  & Dec~$^\mathrm{a}$  & $R_\mathrm{eff}$~$^\mathrm{b}$  & $S_\mathrm{peak}$~$^\mathrm{c}$  &
$S_{850}$~$^\mathrm{d}$  & $M_{850}$~$^\mathrm{e}$  \\
 &  {\sc ID}  & (J2000.0)  & (J2000.0)  & (pc)  & (Jy~beam$^{-1}$)  &
(Jy)  & (M$_\odot$)  \\
\hline
NGC~1333 &  1 & 03:29:11.88 & +31:13:09.4 & 0.020 & 3.47 & 17.56 & 21.02\\
NGC~1333 &  2 & 03:29:03.23 & +31:15:57.5 & 0.030 & 3.34 & 22.23 & 26.61\\
NGC~1333 &  3 & 03:28:55.27 & +31:14:36.5 & 0.043 & 2.76 & 21.71 & 25.98\\
NGC~1333 &  4 & 03:29:10.01 & +31:13:39.4 & 0.024 & 2.18 & 10.71 & 12.82\\
NGC~1333 &  5 & 03:29:01.36 & +31:20:30.5 & 0.052 & 1.40 & 23.48 & 28.11\\
\dots \\
\hline
\end{tabular}
\\
$^\mathrm{a}$~Peak clump flux position. \\
$^\mathrm{b}$~Clump effective radius, the geometric mean of the clump major and minor `sizes' ($\mathrm{\sqrt{Size1\times Size2}}$). Each size is formed from the standard
  deviation of the clump co-ordinates about the centroid position,
  weighted by the pixel data values. \\
$^\mathrm{c}$~Clump peak flux. \\
$^\mathrm{d}$~Clump total flux.\\
$^\mathrm{e}$~Clump mass derived from the total flux assuming optically thin, isothermal dust at 12~K (see section
  \ref{sec:dustmasses}). 
\end{flushleft}
\end{table*}

\begin{table*}
\begin{flushleft}
\caption{Properties of SCUBA clumps found with \gclumps. The full version of this table is available as Supplementary Material to the online version of this article.} 
\label{tab:gclumps_detections}
\begin{tabular}{c c c c c c c c}
\hline
Sub-region~ & Clump~ & RA~$^\mathrm{a}$  & Dec~$^\mathrm{a}$  & $R_\mathrm{eff}$~$^\mathrm{b}$  & $S_\mathrm{peak}$~$^\mathrm{c}$  &
$S_{850}$~$^\mathrm{d}$  & $M_{850}$~$^\mathrm{e}$  \\
 &  {\sc ID}  & (J2000.0)  & (J2000.0)  & (pc)  & (Jy~beam$^{-1}$)  &
(Jy)  & (M$_\odot$)  \\
\hline
NGC~1333 &  1 & 03:29:11.41 & +31:13:21.4 & 0.021 & 3.29 & 18.87 & 22.58\\
NGC~1333 &  2 & 03:29:02.99 & +31:15:54.5 & 0.017 & 3.15 & 12.25 & 14.66\\
NGC~1333 &  3 & 03:28:55.27 & +31:14:36.5 & 0.013 & 2.63 & 6.82 & 8.16\\
NGC~1333 &  4 & 03:29:01.36 & +31:20:33.5 & 0.021 & 1.32 & 9.40 & 11.25\\
NGC~1333 &  5 & 03:29:11.19 & +31:18:27.4 & 0.016 & 1.18 & 4.87 & 5.83\\
\dots\\
\hline
\end{tabular}\\
$^\rmn{a}$~Peak clump flux position. \\
$^\rmn{b}$~Clump effective radius, the geometric mean of the clump
  major and minor `sizes' ($\mathrm{\sqrt{Size1\times Size2}}$). Each size is formed from the standard
  deviation of the clump co-ordinates about the centroid position,
  weighted by the pixel data values. \\
$^\rmn{c}$~Clump peak flux. \\
$^\rmn{d}$~Clump
  total flux.\\ 
$^\rmn{e}$~Clump mass derived from the total flux assuming
  optically thin, isothermal dust at 12~K (see section
  \ref{sec:dustmasses}). 
\end{flushleft}
\end{table*}

\bsp

\label{lastpage}


\begin{thebibliography}{}

\bibitem[\protect\citeauthoryear{{Alves}, {Lombardi} \& {Lada}}{{Alves}
  et~al.}{2007}]{alves07}
{Alves} J.,  {Lombardi} M.,    {Lada} C.~J.,  2007, \aap, 462, L17

\bibitem[\protect\citeauthoryear{{Andr{\'e}}, {Basu} \& {Inutsuka}}{{Andr{\'e}}
  et~al.}{2009}]{andre08}
{Andr{\'e}} P.,  {Basu} S.,    {Inutsuka} S.,  2009, in {Chabrier} G.,  ed.,
  {Structure Formation in Astrophysics}.
Cambridge Univ. Press, Cambridge, p.~254

\bibitem[\protect\citeauthoryear{{Andr{\'e}}, {Belloche}, {Motte} \&
  {Peretto}}{{Andr{\'e}} et~al.}{2007}]{andre07}
{Andr{\'e}} P.,  {Belloche} A.,  {Motte} F.,    {Peretto} N.,  2007, \aap, 472,
  519

\bibitem[\protect\citeauthoryear{{Andr{\'e}} \& {Saraceno}}{{Andr{\'e}} \&
  {Saraceno}}{2005}]{andre05}
{Andr{\'e}} P.,  {Saraceno} P.,  2005, in {Wilson} A.,  ed., Proc. Dusty and
  Molecular Universe: A Prelude to Herschel and ALMA.
Noordwijk, Netherlands: ESA Publ., p.~179

\bibitem[\protect\citeauthoryear{{Andr\'e}, {Ward-Thompson} \&
  {Barsony}}{{Andr\'e} et~al.}{1993}]{andre93}
{Andr\'e} P.,  {Ward-Thompson} D.,    {Barsony} M.,  1993, \apj, 406, 122

\bibitem[\protect\citeauthoryear{{Ballesteros-Paredes} \& {Mac
  Low}}{{Ballesteros-Paredes} \& {Mac Low}}{2002}]{ballesteros02}
{Ballesteros-Paredes} J.,  {Mac Low} M.-M.,  2002, \apj, 570, 734

\bibitem[\protect\citeauthoryear{{Basu} \& {Ciolek}}{{Basu} \&
  {Ciolek}}{2004}]{basu04}
{Basu} S.,  {Ciolek} G.~E.,  2004, \apjl, 607, L39

\bibitem[\protect\citeauthoryear{{Bate} \& {Bonnell}}{{Bate} \&
  {Bonnell}}{2005}]{bate05}
{Bate} M.~R.,  {Bonnell} I.~A.,  2005, \mnras, 356, 1201

\bibitem[\protect\citeauthoryear{{Berry}, {Reinhold}, {Jenness} \&
  {Economou}}{{Berry} et~al.}{2007}]{berry07}
{Berry} D.~S.,  {Reinhold} K.,  {Jenness} T.,    {Economou} F.,  2007, in
  {Shaw} R.~A.,  {Hill} F.,   {Bell} D.~J.,  eds, ASP Conf. Ser. Vol. 376,
  Astronomical Data Analysis Software and Systems XVI.
Astron. Soc. Pac., San Francisco, p.~425

\bibitem[\protect\citeauthoryear{{Blitz}}{{Blitz}}{1993}]{blitz93}
{Blitz} L.,  1993, in {Levy} E.~H.,  {Lunine} J.~I.,  eds, Protostars and
  Planets III.
Univ. Arizona Press, Tucson, p.~125

\bibitem[\protect\citeauthoryear{{Blitz} \& {Shu}}{{Blitz} \&
  {Shu}}{1980}]{blitz80}
{Blitz} L.,  {Shu} F.~H.,  1980, \apj, 238, 148

\bibitem[\protect\citeauthoryear{{Chabrier}}{{Chabrier}}{2005}]{chabrier05}
{Chabrier} G.,  2005, in {Corbelli} E.,  {Palla} F.,   {Zinnecker} H.,  eds,
  The Initial Mass Function 50 Years Later, Astrophys. Space Sci. Library Vol.
  327.
Springer, Dordrecht, p.~41

\bibitem[\protect\citeauthoryear{{Ciolek} \& {Basu}}{{Ciolek} \&
  {Basu}}{2006}]{ciolek06}
{Ciolek} G.~E.,  {Basu} S.,  2006, \apj, 652, 442

\bibitem[\protect\citeauthoryear{{Clark}, {Klessen} \& {Bonnell}}{{Clark}
  et~al.}{2007}]{clark07}
{Clark} P.~C.,  {Klessen} R.~S.,    {Bonnell} I.~A.,  2007, \mnras, 379, 57

\bibitem[\protect\citeauthoryear{{Curtis}, {Richer} \& {Buckle}}{{Curtis}
  et~al.}{2009}]{paper1}
{Curtis} E.~I.,  {Richer} J.~S.,    {Buckle} J.~V.,  2009, \mnras, in press

\bibitem[\protect\citeauthoryear{{Elmegreen} \& {Falgarone}}{{Elmegreen} \&
  {Falgarone}}{1996}]{elmegreen96}
{Elmegreen} B.~G.,  {Falgarone} E.,  1996, \apj, 471, 816

\bibitem[\protect\citeauthoryear{{Elmegreen} \& {Scalo}}{{Elmegreen} \&
  {Scalo}}{2004}]{elmegreen04}
{Elmegreen} B.~G.,  {Scalo} J.,  2004, \araa, 42, 211

\bibitem[\protect\citeauthoryear{{Enoch} et~al.,}{{Enoch}
  et~al.}{2006}]{enoch06}
{Enoch} M.~L.,  et~al., 2006, \apj, 638, 293

\bibitem[\protect\citeauthoryear{{Enoch}, {Evans} II, {Sargent}, {Glenn},
  {Rosolowsky} \& {Myers}}{{Enoch} et~al.}{2008}]{enoch08}
{Enoch} M.~L.,  {Evans} II N.~J.,  {Sargent} A.~I.,  {Glenn} J.,  {Rosolowsky}
  E.,    {Myers} P.,  2008, \apj, 684, 1240

\bibitem[\protect\citeauthoryear{{Gammie}, {Lin}, {Stone} \&
  {Ostriker}}{{Gammie} et~al.}{2003}]{gammie03}
{Gammie} C.~F.,  {Lin} Y.-T.,  {Stone} J.~M.,    {Ostriker} E.~C.,  2003, \apj,
  592, 203

\bibitem[\protect\citeauthoryear{{Goodwin}, {Kroupa}, {Goodman} \&
  {Burkert}}{{Goodwin} et~al.}{2007}]{goodwin07}
{Goodwin} S.~P.,  {Kroupa} P.,  {Goodman} A.,    {Burkert} A.,  2007, in
  {Reipurth} B.,  {Jewitt} D.,   {Keil} K.,  eds, Protostars and Planets V.
Univ. Arizona Press, Tucson, p.~133

\bibitem[\protect\citeauthoryear{{Goodwin}, {Nutter}, {Kroupa}, {Ward-Thompson}
  \& {Whitworth}}{{Goodwin} et~al.}{2008}]{goodwin08}
{Goodwin} S.~P.,  {Nutter} D.,  {Kroupa} P.,  {Ward-Thompson} D.,
  {Whitworth} A.~P.,  2008, \aap, 477, 823

\bibitem[\protect\citeauthoryear{{Goodwin}, {Ward-Thompson} \&
  {Whitworth}}{{Goodwin} et~al.}{2002}]{goodwin02}
{Goodwin} S.~P.,  {Ward-Thompson} D.,    {Whitworth} A.~P.,  2002, \mnras, 330,
  769

\bibitem[\protect\citeauthoryear{{Hatchell} \& {Fuller}}{{Hatchell} \&
  {Fuller}}{2008}]{hatchell08}
{Hatchell} J.,  {Fuller} G.~A.,  2008, \aap, 482, 855

\bibitem[\protect\citeauthoryear{{Hatchell}, {Fuller}, {Richer}, {Harries} \&
  {Ladd}}{{Hatchell} et~al.}{2007}]{hatchell07a}
{Hatchell} J.,  {Fuller} G.~A.,  {Richer} J.~S.,  {Harries} T.~J.,    {Ladd}
  E.~F.,  2007, \aap, 468, 1009 (H07)

\bibitem[\protect\citeauthoryear{{Hatchell}, {Richer}, {Fuller}, {Qualtrough},
  {Ladd} \& {Chandler}}{{Hatchell} et~al.}{2005}]{hatchell05}
{Hatchell} J.,  {Richer} J.~S.,  {Fuller} G.~A.,  {Qualtrough} C.~J.,  {Ladd}
  E.~F.,    {Chandler} C.~J.,  2005, \aap, 440, 151

\bibitem[\protect\citeauthoryear{{Heyer}, {Krawczyk}, {Duval} \&
  {Jackson}}{{Heyer} et~al.}{2009}]{heyer08}
{Heyer} M.,  {Krawczyk} C.,  {Duval} J.,    {Jackson} J.~M.,  2009, \apj, 699,
  1092

\bibitem[\protect\citeauthoryear{{Jijina}, {Myers} \& {Adams}}{{Jijina}
  et~al.}{1999}]{jijina99}
{Jijina} J.,  {Myers} P.~C.,    {Adams} F.~C.,  1999, \apjss, 125, 161

\bibitem[\protect\citeauthoryear{{Johnstone}, {Wilson}, {Moriarty-Schieven},
  {Joncas}, {Smith}, {Gregersen} \& {Fich}}{{Johnstone}
  et~al.}{2000}]{johnstone00b}
{Johnstone} D.,  {Wilson} C.~D.,  {Moriarty-Schieven} G.,  {Joncas} G.,
  {Smith} G.,  {Gregersen} E.,    {Fich} M.,  2000, \apj, 545, 327

\bibitem[\protect\citeauthoryear{{Jones}, {Basu} \& {Dubinski}}{{Jones}
  et~al.}{2001}]{jones01}
{Jones} C.~E.,  {Basu} S.,    {Dubinski} J.,  2001, \apj, 551, 387

\bibitem[\protect\citeauthoryear{{Kirk}, {Johnstone} \& {Di Francesco}}{{Kirk}
  et~al.}{2006}]{kirk06}
{Kirk} H.,  {Johnstone} D.,    {Di Francesco} J.,  2006, \apj, 646, 1009

\bibitem[\protect\citeauthoryear{{Kirk}, {Ward-Thompson} \& {Andr{\'e}}}{{Kirk}
  et~al.}{2005}]{kirkj05}
{Kirk} J.~M.,  {Ward-Thompson} D.,    {Andr{\'e}} P.,  2005, \mnras, 360, 1506

\bibitem[\protect\citeauthoryear{{Klessen} \& {Burkert}}{{Klessen} \&
  {Burkert}}{2000}]{klessen00}
{Klessen} R.~S.,  {Burkert} A.,  2000, \apjss, 128, 287

\bibitem[\protect\citeauthoryear{{Kroupa}}{{Kroupa}}{2001}]{kroupa01}
{Kroupa} P.,  2001, \mnras, 322, 231

\bibitem[\protect\citeauthoryear{{Lada} \& {Lada}}{{Lada} \&
  {Lada}}{2003}]{lada03}
{Lada} C.~J.,  {Lada} E.~A.,  2003, \araa, 41, 57

\bibitem[\protect\citeauthoryear{{Li}, {Norman}, {Mac Low} \& {Heitsch}}{{Li}
  et~al.}{2004}]{li04}
{Li} P.~S.,  {Norman} M.~L.,  {Mac Low} M.-M.,    {Heitsch} F.,  2004, \apj,
  605, 800

\bibitem[\protect\citeauthoryear{{Motte}, {Andre} \& {Neri}}{{Motte}
  et~al.}{1998}]{motte98}
{Motte} F.,  {Andre} P.,    {Neri} R.,  1998, \aap, 336, 150

\bibitem[\protect\citeauthoryear{{Myers} \& {Benson}}{{Myers} \&
  {Benson}}{1983}]{myers83b}
{Myers} P.~C.,  {Benson} P.~J.,  1983, \apj, 266, 309

\bibitem[\protect\citeauthoryear{{Myers}, {Fuller}, {Goodman} \&
  {Benson}}{{Myers} et~al.}{1991}]{myers91}
{Myers} P.~C.,  {Fuller} G.~A.,  {Goodman} A.~A.,    {Benson} P.~J.,  1991,
  \apj, 376, 561

\bibitem[\protect\citeauthoryear{{Nutter} \& {Ward-Thompson}}{{Nutter} \&
  {Ward-Thompson}}{2007}]{nutter07}
{Nutter} D.,  {Ward-Thompson} D.,  2007, \mnras, 374, 1413

\bibitem[\protect\citeauthoryear{{Ossenkopf} \& {Henning}}{{Ossenkopf} \&
  {Henning}}{1994}]{ossenkopf94}
{Ossenkopf} V.,  {Henning} T.,  1994, \aap, 291, 943

\bibitem[\protect\citeauthoryear{{Padoan} \& {Nordlund}}{{Padoan} \&
  {Nordlund}}{2002}]{padoan02}
{Padoan} P.,  {Nordlund} {\AA}.,  2002, \apj, 576, 870

\bibitem[\protect\citeauthoryear{{Reid} \& {Wilson}}{{Reid} \&
  {Wilson}}{2006}]{reid06}
{Reid} M.~A.,  {Wilson} C.~D.,  2006, \apj, 650, 970

\bibitem[\protect\citeauthoryear{{Rosolowsky}, {Pineda}, {Foster}, {Borkin},
  {Kauffmann}, {Caselli}, {Myers} \& {Goodman}}{{Rosolowsky}
  et~al.}{2008}]{rosolowsky08}
{Rosolowsky} E.~W.,  {Pineda} J.~E.,  {Foster} J.~B.,  {Borkin} M.~A.,
  {Kauffmann} J.,  {Caselli} P.,  {Myers} P.~C.,    {Goodman} A.~A.,  2008,
  \apjss, 175, 509

\bibitem[\protect\citeauthoryear{{Salpeter}}{{Salpeter}}{1955}]{salpeter55}
{Salpeter} E.~E.,  1955, \apj, 121, 161

\bibitem[\protect\citeauthoryear{{Scalo}}{{Scalo}}{2005}]{scalo05}
{Scalo} J.,  2005, in {Corbelli} E.,  {Palla} F.,   {Zinnecker} H.,  eds, The
  Initial Mass Function 50 Years Later, Astrophys. Space Sci. Library Vol. 327.
Springer, Dordrecht, p.~23

\bibitem[\protect\citeauthoryear{{Schnee}, {Li}, {Goodman} \&
  {Sargent}}{{Schnee} et~al.}{2008}]{schnee08b}
{Schnee} S.,  {Li} J.,  {Goodman} A.~A.,    {Sargent} A.~I.,  2008, \apj, 684,
  1228

\bibitem[\protect\citeauthoryear{{Schneider} \& {Brooks}}{{Schneider} \&
  {Brooks}}{2004}]{schneider04}
{Schneider} N.,  {Brooks} K.,  2004, Publ. Astron. Soc. Australia, 21, 290

\bibitem[\protect\citeauthoryear{{Shirley}, {Evans} \& {Rawlings}}{{Shirley}
  et~al.}{2002}]{shirley02}
{Shirley} Y.~L.,  {Evans} II N.~J.,    {Rawlings} J.~M.~C.,  2002, \apj, 575,
  337

\bibitem[\protect\citeauthoryear{{Simpson}, {Nutter} \&
  {Ward-Thompson}}{{Simpson} et~al.}{2008}]{simpson08}
{Simpson} R.~J.,  {Nutter} D.,    {Ward-Thompson} D.,  2008, \mnras, 391, 205

\bibitem[\protect\citeauthoryear{{Smith}, {Clark} \& {Bonnell}}{{Smith}
  et~al.}{2008}]{smith_clark08}
{Smith} R.~J.,  {Clark} P.~C.,    {Bonnell} I.~A.,  2008, \mnras, 391, 1091

\bibitem[\protect\citeauthoryear{{Solomon}, {Rivolo}, {Barrett} \&
  {Yahil}}{{Solomon} et~al.}{1987}]{solomon87}
{Solomon} P.~M.,  {Rivolo} A.~R.,  {Barrett} J.,    {Yahil} A.,  1987, \apj,
  319, 730

\bibitem[\protect\citeauthoryear{{Stutzki}, {Bensch}, {Heithausen}, {Ossenkopf}
  \& {Zielinsky}}{{Stutzki} et~al.}{1998}]{stutzki98}
{Stutzki} J.,  {Bensch} F.,  {Heithausen} A.,  {Ossenkopf} V.,    {Zielinsky}
  M.,  1998, \aap, 336, 697

\bibitem[\protect\citeauthoryear{{Stutzki} \& {G\"usten}}{{Stutzki} \&
  {G\"usten}}{1990}]{stutzki90}
{Stutzki} J.,  {G\"usten} R.,  1990, \apj, 356, 513

\bibitem[\protect\citeauthoryear{{Swift} \& {Williams}}{{Swift} \&
  {Williams}}{2008}]{swift08}
{Swift} J.~J.,  {Williams} J.~P.,  2008, \apj, 679, 552

\bibitem[\protect\citeauthoryear{{Tassis}}{{Tassis}}{2007}]{tassis07}
{Tassis} K.,  2007, \mnras, 379, L50

\bibitem[\protect\citeauthoryear{{Tassis}, {Dowell}, {Hildebrand}, {Kirby} \&
  {Vaillancourt}}{{Tassis} et~al.}{2009}]{tassis09}
{Tassis} K.,  {Dowell} C.~D.,  {Hildebrand} R.~H.,  {Kirby} L.,
  {Vaillancourt} J.~E., 2009, \mnras, in press

\bibitem[\protect\citeauthoryear{{Ward-Thompson}, {Scott}, {Hills} \&
  {Andr\'e}}{{Ward-Thompson}
  et~al.}{1994}]{wardthompson94}
{Ward-Thompson} D.,  {Scott} P.~F., {Hills} R.~E., {Andr{\'e}} P.,
  1994, \mnras, 268, 276

\bibitem[\protect\citeauthoryear{{Ward-Thompson}, {Andr{\'e}}, {Crutcher},
  {Johnstone}, {Onishi} \& {Wilson}}{{Ward-Thompson}
  et~al.}{2007}a]{wardthompson07}
{Ward-Thompson} D.,  {Andr{\'e}} P.,  {Crutcher} R.,  {Johnstone} D.,  {Onishi}
  T.,    {Wilson} C.,  2007a, in {Reipurth} B.,  {Jewitt} D.,   {Keil} K.,  eds,
  Protostars and Planets V.
Univ. Arizona Press, Tucson, p.~33

\bibitem[\protect\citeauthoryear{{Ward-Thompson} et~al.,}{{Ward-Thompson}
  et~al.}{2007}b]{gbs}
{Ward-Thompson} D.,  et~al., 2007b, PASP, 119, 855

\bibitem[\protect\citeauthoryear{{Williams}, {Blitz} \& {McKee}}{{Williams}
  et~al.}{2000}]{williams00}
{Williams} J.~P.,  {Blitz} L.,    {McKee} C.~F.,  2000, in {Mannings} V.,
  {Boss} A.~P.,   {Russell} S.~S.,  eds, Protostars and Planets IV.
Univ. Arizona Press, Tucson, p.~97

\bibitem[\protect\citeauthoryear{{Williams}, {de Geus} \& {Blitz}}{{Williams}
  et~al.}{1994}]{williams94}
{Williams} J.~P.,  {de Geus} E.~J.,    {Blitz} L.,  1994, \apj, 428, 693

\bibitem[\protect\citeauthoryear{{Young}, {Shirley}, {Evans} II \&
  {Rawlings}}{{Young} et~al.}{2003}]{young03}
{Young} C.~H.,  {Shirley} Y.~L.,  {Evans} II N.~J.,    {Rawlings} J.~M.~C.,
  2003, \apjss, 145, 111

\bibitem[\protect\citeauthoryear{{Zel'Dovich}}{{Zel'Dovich}}{1970}]{zeldovich7%
0}
{Zel'Dovich} Y.~B.,  1970, \aap, 5, 84

\end{thebibliography}
\end{document}